\documentclass[12pt]{article}

\usepackage[utf8x]{inputenc}
\usepackage{amsfonts}
\usepackage[]{hyperref}
\hypersetup{linktocpage}
\usepackage{mathtools}
\usepackage{amsmath,amssymb,latexsym,cite}
\usepackage{axodraw}

\begin{document}

\begin{center}

\vspace*{0.5in}

{\large\bf More Toda-like (0,2) mirrors}

\vspace{0.2in}

Zhuo Chen, Jirui Guo, Eric Sharpe, Ruoxu Wu

\vspace*{0.2in}

\begin{tabular}{c}
Physics Department \\
Robeson Hall (0435) \\
Virginia Tech \\
Blacksburg, VA  24061 
\end{tabular}

{\tt zhuo2012@vt.edu}, {\tt jrkwok@vt.edu}, {\tt ersharpe@vt.edu}, {\tt ronwu@vt.edu}

$\,$

\end{center}

In this paper, we extend our previous work to construct $(0,2)$ 
Toda-like mirrors to $A/2$-twisted theories on 
more general spaces, as part of a program of understanding (0,2) mirror
symmetry.  Specifically, 
we propose $(0,2)$ mirrors to GLSMs on toric del Pezzo surfaces and
Hirzebruch surfaces 
with deformations of the tangent bundle. 
We check the results by comparing correlation functions, global symmetries,
as well as
geometric blowdowns with the corresponding $(0,2)$ Toda-like mirrors. 
We also briefly discuss Grassmannian manifolds.

\begin{flushleft}
May 2017
\end{flushleft}

\newpage

\tableofcontents

\newpage

\section{Introduction}

Mirror symmetry has historically
been of great interest to both physicists and 
mathematicians. In heterotic string compactifications, there is a natural
generalization, known as $(0,2)$ mirror symmetry, see
{\it e.g.} \cite{bsw,blum-sethi,Adams:2003zy,mp,Melnikov:2012hk}.
For mathematics, $(0,2)$ mirror symmetry yields quantum sheaf cohomology,
a generalization of quantum  cohomology in $(2,2)$ theories,
see {\it e.g.} 
\cite{Katz:2004nn, Sharpe:2006qd, Guffin:2007mp, Melnikov:2007xi, McOrist:2007kp, McOrist:2008ji, ade, Guffin:2011mx, Donagi:2011uz, Donagi:2011va, 
Melnikov:2012hk, Closset:2015ohf, Guo:2015caf, Guo:2016suk}. 

A perturbative heterotic compactification is defined by a worldsheet
theory with (0,2) supersymmetry.
A $(0,2)$ nonlinear sigma model is defined by a pair $(X, \mathcal{E})$,
with $X$ a K\"ahler manifold and 
$\mathcal{E} \rightarrow X$
a holomorphic vector bundle, satisfying Green-Schwarz anomaly cancellation
\begin{displaymath}
\text{ch}_2 (\mathcal{E}) = \text{ch}_2 (TX).
\end{displaymath}
In cases in which $X$ is Calabi-Yau, so that the nonlinear sigma model
above flows to a SCFT, (0,2) mirror symmetry states that
there is a dual pair  
$(X', \mathcal{E}')$ which gives rise to the same SCFT. 
If $X$ is Fano, then the (0,2) mirror will be a (0,2) Landau-Ginzburg model.

In this paper, we will focus on topological twists of these theories.
In (0,2) theories, broadly speaking, two topological twists exist, now known
as the A/2 and B/2 twists.  In the case of the nonlinear sigma models above,
the A/2 twist will exist when
$\det \mathcal{E}^* \cong K_X$,
and the B/2 twist will exist when
$\det \mathcal{E} \cong K_X$.  Clearly, both the Green-Schwarz condition
and the conditions for the twists will be satisfied when one takes
$\mathcal{E} = TX$, in which case, the $A/2$ theory becomes the ordinary
$A$ model topological field theory, and the $B/2$ theory becomes the
ordinary $B$ model topological field theory.

Quantum sheaf cohomology emerges as the OPE algebra of the A/2-twisted
theory, forming a precise $(0,2)$ analogue of ordinary quantum cohomology. 
To be specific, recall that
the ordinary quantum cohomology ring is a ring of local operators defined 
in the $A$ twist of a nonlinear sigma model on $X$ as 
BRST-closed states of the form
\begin{equation*}
b_{i_1 \cdots i_p \bar{\imath}_1 \cdots \bar{\imath}_q} \chi^{i_1} \cdots \chi^{i_p} \chi^{\bar{\imath}_1} \cdots \chi^{\bar{\imath}_q}.
\end{equation*}
These BRST-closed states can be identified with closed differential forms 
on the target space $X$, elements of $H^q(X, \Omega^p) = H^{p,q}(X)$. 
Similarly, the quantum sheaf cohomology ring is a ring of local operators 
defined in the $A/2$ twist of a nonlinear sigma model on $X$ as
right-BRST-closed states of the form
\begin{equation*}
b_{a_1 \cdots a_p \overline{\imath}_1 \cdots \overline{\imath}_q} 
\lambda_-^{a_1} \cdots \lambda_-^{a_p}
\psi_+^{\overline{\imath}_1} \cdots \psi_+^{\overline{\imath}_q}.
\end{equation*}
These right-BRST-closed states can be identified with 
$\overline{\partial}$-closed bundle-valued differential forms, 
elements of $H^q (X, \wedge^p \mathcal{E}^*)$.

Ordinary mirror symmetry exchanges $A$ twists with $B$ twists, 
which means an $A$ twisted nonlinear sigma model is equivalent to a $B$ twisted nonlinear sigma model on the mirror Calabi-Yau manifold \cite{Witten:1991zz}. 
Similarly, $(0,2)$ mirror symmetry exchanges $A/2$ twists with $B/2$ twists, meaning the $A/2$ twisted nonlinear sigma model on $(X,\mathcal{E})$ is equivalent to the $B/2$ twisted nonlinear sigma model on the mirror $(X', \mathcal{E}')$. 

A version of mirror symmetry also exists for Fano spaces
(see {\it e.g.} \cite{giv1,ehy} for a few early references). An $A$ twisted 
(respectively $A/2$ twisted) nonlinear sigma model on a Fano manifold $X$ 
is equivalent to a $B$ twist (respectively $B/2$ twist) of 
a $(0,2)$ Landau-Ginzburg model. 
The $(2,2)$ version of this duality is well studied and is given by Toda duals 
to Fano manifolds \cite{Hori:2000kt}. Comparatively, little is known about 
$(0,2)$ analogues. Our previous work \cite{Chen:2016tdd} constructed $(0,2)$ 
Toda-like mirrors to products of projective spaces, generalizing the only 
example previously in the literature \cite{Adams:2003zy}. 
The goal of this paper is to extend the construction of (0,2) Landau-Ginzburg 
mirrors to more interesting geometries, as deformations of (2,2) 
Landau-Ginzburg mirrors, to help pave the way for a more systematic 
understanding of (0,2) mirror symmetry.
Other recent work on two-dimensional (0,2) theories from different
directions includes {\it e.g.} \cite{Schafer-Nameki:2016cfr,Franco1,gp1,Franco:2016qxh, Franco:2016fxm,Lawrie:2016rqe,
Franco:2017cjj,ahhm,ahhm2,Dedushenko:2017tdw}.

In this paper, we extend our previous work \cite{Chen:2016tdd} and explore 
$(0,2)$ Toda-like mirrors to $A/2$-twisted theories on more spaces. 
Our previous work studied the $(0,2)$ Toda-like mirror to $A/2$ model 
on $\mathbb{P}^n \times \mathbb{P}^m$.  In this paper, we will construct
ansatzes for (0,2) mirrors to toric del Pezzo surfaces and Hirzebruch\footnote{
More precisely, as we will explain later, gauged linear sigma models (GLSMs)
\cite{Witten:1993yc} for Hirzebruch surfaces.
Most Hirzebruch surfaces are not Fano, and so the UV limits of their GLSMs
are not Hirzebruch surfaces but rather different geometries, so in principle
we are actually describing mirrors to those different surfaces.
}
surfaces, as part of an on-going program to understand (0,2) mirror symmetry.
These ansatzes will be tested in several different ways:
\begin{itemize}
\item First, each case reduces to an ordinary (2,2) mirror along
the (2,2) locus.  
\item We check that the fields in the Landau-Ginzburg
vacua obey the quantum sheaf cohomology relations of their $A/2$-model
partners.
\item We check in each case that all genus zero correlation functions
of the proposed $B/2$-twisted Landau-Ginzburg mirror match those of the
original $A/2$-twisted (0,2) theory.   
\item Amongst the toric del Pezzo mirrors, we check that our proposed
mirrors are related by blowdowns as dictated by\footnote{
As by definition the del Pezzos are Fano, the UV GLSM phases correspond to
the naive geometries.
} geometry.
\item As an implicit check, we also give a proposal for (0,2) mirrors
of Hirzebruch surfaces\footnote{
For degree greater than one, Hirzebruch surfaces are not Fano; nevertheless,
one expects their sigma models to have isolated vacua in the IR, hence
a Toda-type mirror is expected.
} of arbitrary degree, which not only correctly
captures the genus zero correlation functions, but also includes as
special cases our
previous proposed mirror for ${\mathbb P}^1 \times {\mathbb P}^1$
\cite{Chen:2016tdd} and for the del Pezzo $dP_1$ above, thereby
demonstrating that the ${\mathbb P}^1 \times {\mathbb P}^1$ and $dP_1$
mirrors are indeed elements of a sequence of mirrors, as one would expect.
\end{itemize}

There is another subtlety we shall encounter in the form of the $J$ functions
defining the (0,2) superpotential.  Specifically, they will sometimes have
poles away from the origin.  Now, ordinary (2,2) mirrors to projective
spaces and Fano varieties will often have superpotential terms proportional
to $1/X^n$ for $n>0$, but it is understood that those Landau-Ginzburg models
are defined over algebraic tori of the form $({\mathbb C}^{\times})^k$,
so that the target space does not include places where $X = 0$,
and hence the theory never encounters a divergent superpotential.  By contrast,
in this paper we will encounter some examples which have poles at points
which are not disallowed.  As a result, we interpret these theories in
a low-energy effective theory sense -- so long as no vacua are located
at those poles, we can understand the theory in a neighborhood of the vacua,
which excludes the poles.  (Similar remarks have been applied to understand
GLSMs for generalized Calabi-Yau complete intersections 
\cite{aaggl,bh1611,agcl}.)  Of course, this also means that these
theories are not UV-complete, but we will leave searches for UV-complete
descriptions for other work.

We begin in section~\ref{sect:review} by reviewing the construction of
\cite{Hori:2000kt} of (2,2) mirrors (Toda duals) to Fano spaces realized
in GLSMs, as well as previous results of \cite{Chen:2016tdd} on
(0,2) mirrors to products of projective spaces with deformations of the
tangent bundle, which form the heart of both the proposed del Pezzo and
Hirzebruch (0,2) mirrors.  In section~\ref{sec: dp} we turn our attention to
toric del Pezzo surfaces, giving proposed mirrors to toric del Pezzo surfaces
with tangent bundle deformations, checking that correlation functions match
as well as that mirrors to blowdowns are related in the fashion one would 
expect.  In section~\ref{sec: hirzebruch} we turn to Hirzebruch surfaces,
and give a proposal that generalizes our results for
${\mathbb F}_1=dP_1$ and ${\mathbb F}_0 = {\mathbb P}^1 \times {\mathbb P}^1$,
and also satisfies consistency tests. 
Finally in section~\ref{sec: grassmannian} we briefly discuss a possible
(0,2) analogue of the proposed Grassmannian mirror in 
\cite{Hori:2000kt}[appendix A].  An appendix discusses quantum cohomology
of $dP_1$, which figures into its mirror.

\section{Review}
\label{sect:review}

\subsection{Review of (2,2) Toda dual theories}
\label{sect:review:22}

Consider a (2,2) supersymmetric abelian GLSM, with gauge group 
$U(1)^k$ and $n$ chiral superfields $\Phi_i$.
Let $Q_i^a$ denote the
charge of the $i$th chiral superfield under the $a$th factor in
the gauge group. Following \cite{Hori:2000kt}, 
the mirror\footnote{
If the toric variety is Fano, this will yield the mirror of the Fano
phase.  Otherwise, it may yield the mirror of a different phase.
} of an A-twisted theory of this form is a 
Landau-Ginzburg model with a superpotential of the form
\begin{equation}   \label{eq:22mirror-w}
W \: = \: \sum_{a=1}^k \Sigma_a \left( \sum_{i=1}^n Q_i^a Y_i - r_a
\right) \: + \: \sum_{i=1}^n \exp(Y_i),
\end{equation}
where the $Y_i$ are twisted chiral superfields in one-to-one correspondence
with chiral superfields in the original theory.  We integrate out the 
$\Sigma_a$'s to recover the usual form.

It will also be useful to track R-symmetries, as a consistency test on
our proposals.
Recall that a two-dimensional $\mathcal{N}=(2,2)$ theory has classical
left-moving $U(1)_L$ and a right-moving $U(1)_R$ R-symmetries,
\begin{align*}
U(1)_R:& \quad \theta^+ \mapsto e^{-i \kappa} \theta^+, \\
U(1)_L:& \quad \theta^- \mapsto e^{-i \kappa} \theta^-.
\end{align*}

Denoting the generators of the R-symmetry $U(1)_L \times U(1)_R$ 
as $J_L$ and $J_R$ respectively, then one can combine them to get the 
vector R-symmetry $U(1)_V$ and axial R-symmetry $U(1)_A$ with
generators 
\begin{equation*}
J_V =  \frac{1}{2} (J_R + J_L), \quad J_A =  \frac{1}{2} ( J_R - J_L).
\end{equation*}
Chiral superfields transform under the R-symmetries as follows
\cite{Hori:2000kt}[equ'ns (2.11)-(2.12)],
\begin{align*}
R_V \Phi_i ( x, \theta^{\pm}, \bar{\theta}^{\pm} ) &= 
e^{- i \alpha q_V^i} \Phi_i ( x, e^{ -i \alpha} \theta^{\pm}, e^{ i \alpha } \bar{\theta}^{\pm} ) , \\
R_A \Phi_i ( x, \theta^{\pm}, \bar{\theta}^{\pm} ) &= 
e^{- i \beta q_A^i} \Phi_i ( x, e^{ \mp i \beta} \theta^{\pm}, 
e^{ \pm i \beta } \bar{\theta}^{\pm} ),
\end{align*}
where $q_{V, A}^i$ denote the vector and axial R-charges of $\Phi_i$,
chosen so that the superpotential has vector charge $2$ and
axial charge $0$.
(A twisted superpotential, a function of twisted chiral superfields,
has vector charge $0$ and axial charge $2$.)
In components,
\begin{align*}
R_V: & \: \:  x \mapsto e^{i \alpha q_V} x, \: \: \:
\psi_{\pm} \mapsto e^{i \alpha (q_V + 1)} \psi_{\pm}, \\
R_A: & \: \:  x \mapsto e^{i \beta q_A} x, \: \: \:
\psi_{\pm} \mapsto e^{i \beta (q_A \pm 1)} \psi_{\pm}.
\end{align*}
In the quantum theory, the axial R symmetry is typically anomalous.

Now we turn to the mirror theory.
We assume the original theory has no superpotential (as we are taking
the mirror of a toric variety), so the vector and axial R-charges of the 
original chiral superfields both
vanish.
The twisted chiral superfields $Y_i$ transform as\footnote{
We follow the conventions of \cite{Hori:2000kt} in
using the same `axial,' `vector' terminology to describe both the
original symmetry and its mirror, to assist in tracking the symmetries.
} \cite{Hori:2000kt}[equ'ns (3.29)-(3.30)]
\begin{align}
R_V Y_i ( x, \theta^{\pm}, \bar{\theta}^{\pm} ) &= 
Y_i ( x, e^{ -i \alpha} \theta^{\pm}, e^{ i \alpha } \bar{\theta}^{\pm} ) ,
\label{eq:r-vector-y} \\
R_A  Y_i ( x, \theta^{\pm}, \bar{\theta}^{\pm} ) &= 
Y_i ( x, e^{ \mp i \beta} \theta^{\pm}, e^{ \pm i \beta } \bar{\theta}^{\pm} )
 - 2i \beta.
\label{eq:r-axial-y}
\end{align}
It is straightforward to see that in the mirror Landau-Ginzburg model
defined by~(\ref{eq:22mirror-w}), the vector R-symmetry is unbroken, but the
axial R-symmetry is broken classically by the superpotential, 
corresponding to the fact that
in the original theory, the axial R-symmetry is anomalous.

For example, consider
the mirror to ${\mathbb P}^n$,
which (after integrating out $\Sigma$) is a Landau-Ginzburg theory
defined by the (twisted) superpotential
\begin{equation*}
\int {\rm d} \theta^+ \, {\rm d} \bar{\theta}^- 
\widetilde{W} + c.c. 
= \int {\rm d} \theta^+ \, {\rm d} \bar{\theta}^- \left( \sum_{i=1}^n X_i + \frac{q}{ \prod_{i=1}^n X_i} \right) + c.c. ,
\end{equation*}
where $X_i = \exp{Y_i}$ and $q = \exp (-r)$.  
Here, the last term, $q \prod X_i^{-1}$, classically breaks the 
axial R symmetry unless 
\begin{displaymath}
\exp( 2 i \beta (n+1) ) \: = \: 1,
\end{displaymath}
corresponding to the anomaly of the original theory, breaking the
original $U(1)$ symmetry to a ${\mathbb Z}_{2(n+1)}$ subgroup.

\subsection{(0,2) mirrors to products of projective spaces}
\label{sect:rev02-products}

Comparatively little is known about analogous (0,2) mirrors to
non-Calabi-Yau spaces.  As an attempt to
rectify this situation, the recent paper \cite{Chen:2016tdd} constructed
and checked ansatzes for (0,2) mirrors to products of projective spaces
with deformations of the tangent bundle.  In particular, this paper will
make use of results for ${\mathbb P}^1 \times {\mathbb P}^1$, which as
both a Hirzebruch surface and a toric Fano surface, will be a starting
point for several discussions in this paper.

To make this paper self-contained, we briefly review the pertinent 
results here.

A deformation of the tangent bundle of ${\mathbb P}^1 \times
{\mathbb P}^1$ is defined as the cokernel ${\cal E}$ below:
\begin{displaymath}
0 \: \longrightarrow \: {\cal O}^2 \: \stackrel{E}{\longrightarrow} \:
{\cal O}(1,0)^2 \oplus {\cal O}(0,1)^2 \: \longrightarrow \: {\cal E} \:
\longrightarrow \: 0,
\end{displaymath}
where $E$ is the map
\begin{displaymath}
E \: = \: \left[ \begin{array}{cc}
A x & B x \\
C \tilde{x} & D \tilde{x} \end{array} \right],
\end{displaymath}
where $x$, $\tilde{x}$ are two-component vectors of homogeneous coordinates
on either ${\mathbb P}^1$ factor and $A$, $B$, $C$, $D$ are four
constant $2 \times 2$ matrices, whose parameters define the deformation.

The proposed mirror \cite{Chen:2016tdd} is a (0,2) Landau-Ginzburg
model defined in superspace by the (0,2) superpotential
\begin{align*}
\int {\mathrm d} \theta^+ W=& \int {\mathrm d} \theta^+ \left(
\Lambda J+\tilde{\Lambda} \tilde{J} \right),
\end{align*}
where $\Lambda$ and $\tilde{\Lambda}$ are Fermi superfields, 
$X_i = \exp ( Y_i)$, 
$J$ and $\tilde{J}$ are holomorphic functions given here by
\begin{eqnarray}
J & = & aX_1+b\frac{X_2^2}{X_1}+
\mu X_2-\frac{q_1}{X_1}, \label{eq:p1p1_1} \\
\tilde{J} & = & d X_2+c\frac{X_1^2}{X_2}+\nu X_1-\frac{q_2}{X_2},
\label{eq:p1p1_2}
\end{eqnarray}
and 
\begin{displaymath}
a = \det A, \: \: \:
b = \det B, \: \: \:
c = \det C, \: \: \: 
d = \det D, 
\end{displaymath}
\begin{eqnarray*}
\mu & = & \det(A+B) - \det A - \det B, \\
\nu & = & \det(C+D) - \det C - \det D.
\end{eqnarray*}
For readers not acquainted with (0,2) theories, on the (2,2) locus the
$J$'s become derivatives of the (2,2) superpotential (with respect to the
(2,2) chiral multiplets of which the (0,2) Fermi multiplets are half).
It is straightforward to check that, indeed, in this case along the (2,2)
locus, $J = \partial W / \partial Y_1$, $\tilde{J} = \partial W/\partial Y_2$
for
\begin{displaymath}
W \: = \: X_1 \: + \: \frac{q_1}{X_1} \: + \: X_2 \: + \: \frac{q_2}{X_2},
\end{displaymath}
the (2,2) Toda dual to ${\mathbb P}^1 \times {\mathbb P}^1$.

Next, let us consider $U(1)$ symmetries.  On the (2,2) locus, we have
both\footnote{
In fact, because the target is a product of two spaces, on the (2,2) locus
we have additional symmetries obtained by acting nontrivially on fields
associated with only a single ${\mathbb P}^1$.  A generic (0,2) deformation
breaks such symmetries; only those (2,2)-locus-symmetries acting 
symmetrically on both ${\mathbb P}^1$ factors survive, and so we focus
on those here.
} left- and right-moving R-symmetries; in (0,2), we have instead
a right-moving R symmetry and a left-moving $U(1)$ symmetry which becomes
an R-symmetry on the (2,2) locus.  We can combine those two chiral
actions in symmetric and antisymmetric combinations to form vector and
axial symmetries $U(1)_{V, A}$
which become the vector and axial R-symmetries $R_{V, A}$ on the
(2,2) locus.  Explicitly, on (0,2) chiral and Fermi multiplets $Y_i$,
$\Lambda^i$, respectively:
\begin{align*}
U(1)_V  Y_i(x, \theta^+, \bar{\theta}^+) &= 
Y_i(x, e^{-i \alpha} \theta^+, e^{i \alpha} \bar{\theta}^+), \\
U(1)_V   \Lambda^i(x, \theta^+, \bar{\theta}^+) &=
e^{- i \alpha}\Lambda(x,e^{-i \alpha} \theta^+, e^{i \alpha} \bar{\theta}^+), \\%
\, & \\
U(1)_A Y_i(x, \theta^+, \bar{\theta}^+) &=
Y_i(x, e^{- i \beta} \theta^+, e^{i \beta} \bar{\theta}^+) - 2 i \beta, \\
U(1)_A \Lambda^i(x, \theta^+, \bar{\theta}^+) &=
e^{i \beta} \Lambda^i(x, e^{- i \beta} \theta^+, e^{i \beta} \bar{\theta}^+).
\end{align*}

It is straightforward to check that the proposed (0,2) mirror above is
invariant under $U(1)_V$, but not under $U(1)_A$ because of the $q_1/X_1$,
$q_2/X_2$ terms, except for a finite subgroup defined by
$\exp(4i\beta) = 1$.  This matches results for the $A/2$ model on
${\mathbb P}^1 \times {\mathbb P}^1$, which is invariant under the
vector $U(1)$ but the axial $U(1)$ is anomalous and so is broken to a finite
subgroup.  In fact, this matches results for the (2,2) locus of the $A/2$
model -- since anomalies are computed by indices, they are invariant under
deformations, and so as a matter of principle one should obtain the same
results for the (2,2) locus as its (0,2) deformations.

\section{Del Pezzo surfaces} \label{sec: dp}

In this section, we will discuss mirrors to toric del Pezzo surfaces.
We will use the notation $dP_k$ to indicate ${\mathbb P}^2$ blown up at $k$
points.

\subsection{The first del Pezzo surface, $dP_1$}

The first del Pezzo surface we will consider,
$dP_1$, corresponding to a single blowup of
${\mathbb P}^2$, is isomorphic to the first Hirzebruch surface $\mathbb{F}_1$. 
As mirrors to 
higher del Pezzo surfaces will be constructed on the `foundation' of
$dP_1$, let us very by describing its (2,2) and (0,2) mirrors.
(Appendix~\ref{app:qc:dp1} reviews some standard results on quantum
cohomology of $dP_1$, standard in the math community but perhaps less
well-known in the physics community, that are pertinent for the mirror.)

\subsubsection{(2,2) and proposed (0,2) mirrors}
\label{sect:dp1:02mirror}

The del Pezzo surface $dP_1$
can be described as a toric variety by a fan with edges 
$(1,0)$, $(0,1)$, $(-1,-1)$, $(0,-1)$. A corresponding GLSM is defined by
four chiral superfields $\phi_i, i= 1 \dots 4$ charged under the
gauge group $U(1) \times U(1)$ as follows:
\[
\begin{tabular}{ c c c c}
(1,0) & (-1,-1) & (0,1) & (0,-1) \\
\hline
1 & 1 & 1 & 0 \\
0 & 0 & 1 & 1 \\
\end{tabular}
\]
The quantum cohomology relations\footnote{Note that example~7.3 in
\cite{Miranda:1994} gives the same quantum cohomology ring relations after
identifying $\psi \sim f$, $\tilde{\psi} \sim e$,
$q_1 \sim r$, $q_1 q_2^{-1} \sim q$.} are
\begin{align*}
\psi^2 ( \psi + \tilde{\psi}) = q_1, \\
(\psi + \tilde{\psi}) \tilde{\psi} = q_2. \\
\end{align*}

As reviewed in section~\ref{sect:review:22},
the (2,2) mirror to a sigma model on $dP_1 = {\mathbb F}_1$ is 
\cite{Hori:2000kt} a
Landau-Ginzburg theory with superpotential
\begin{displaymath}
W = \exp ( Y_1) + \exp ( Y_2) + \exp ( Y_3) + \exp (  Y_4),
\end{displaymath}
where the fields obey the constraints
\begin{displaymath}
Y_1 + Y_2 + Y_3 = r_1, \quad Y_3 + Y_4 = r_2.
\end{displaymath}
We will describe
ansatzes for (0,2) mirrors based on two different solutions of the
constraints above.

Our first description of the (2,2) $B$-twisted mirror to the $A$-twisted
theory is written in terms of $Y_1$ and $Y_3$.
Define
$X_1=\exp ( Y_1)$ and $X_3=\exp ( Y_3)$,
then the mirror can be described as a Landau-Ginzburg model
over $({\mathbb C}^{\times})^2$ with superpotential
\begin{equation}
W = X_1 + X_3 + \frac{q_2}{X_3} + \frac{q_1}{X_1 X_3}. \label{eq:22_sp_one}
\end{equation}
(This matches the mirror given in \cite{Hori:2000kt}[equ'n (5.19)].)

An alternative description of the mirror to the same theory is written in terms of
$Y_1$ and $Y_4$.
Define $X_1 = \exp (  Y_1)$ and $X_4 = \exp (  Y_4)$, then
on the $(2,2)$ locus, the mirror superpotential is 
\begin{equation}
W = X_1 + X_4 + \frac{q_1}{q_2} \frac{X_4}{X_1} + \frac{q_2}{X_4}. 
\label{eq:dp1-22-2}
\end{equation}
On the (2,2) locus, this can be related to the previous expression via
the field redefinition
\begin{displaymath}
X_4 \: = \: \frac{q_2}{X_3}.
\end{displaymath}
(Analogous field redefinitions can be computed to relate the (0,2) mirrors
we discuss next, but their expressions for general parameters are both
extremely unwieldy and unhelpful, so we omit them from this paper.)

The $(0,2)$ deformations of $dP_1$ 
are defined by a pair of $2 \times 2$ matrices $A,B$,
and complex numbers $\gamma_1, \gamma_2, \alpha_1, \alpha_2$,
that define a deformation $\mathcal{E}$ of the tangent bundle
\[
0 \: \longrightarrow \: {\cal O}^{\oplus 2} \: \xrightarrow{E} \:
 {\cal O}(1,0)^{\oplus 2} \oplus {\cal O}(1,1) \oplus {\cal O}(0,1)
 \: \longrightarrow {\cal E} \: \longrightarrow 0 ,
\]
where $E$ is
\begin{displaymath}
E = 
\begin{bmatrix}
Ax & Bx \\
\gamma_1 s & \gamma_2 s \\
\alpha_1 t & \alpha_2 t
\end{bmatrix},
\end{displaymath}
with
\begin{displaymath}
x= 
\begin{bmatrix}
u \\ 
v
\end{bmatrix}.
\end{displaymath}
The $(2,2)$ locus is given by the special case
\begin{displaymath}
A = I, \quad B = 0, \quad \gamma_1 = 1, \quad \gamma_2 = 1, \quad \alpha_1 = 0, \quad \alpha_2 = 1.
\end{displaymath}
If we define
\begin{displaymath}
Q_{(k)} = \det (\psi A + \tilde{\psi} B), \quad Q_{(s)} = \psi \gamma_1 + \tilde{\psi} \gamma_2, \quad Q_{(t)} = \psi \alpha_1 + \tilde{\psi} \alpha_2,
\end{displaymath}
then the quantum sheaf cohomology ring relations are given by \cite{Melnikov:2012hk}
\begin{equation}
Q_{(k)} Q_{(s)} = q_1, \quad Q_{(s)} Q_{(t)} = q_2.
\label{eq: 1_qsc2}
\end{equation}

Next, we shall give an ansatz for a ($B/2$-twisted)
(0,2) Landau-Ginzburg theory which is
mirror to the $A/2$ model on $dP_1$ with
deformed tangent bundle as above.  For readers not familiar with (0,2)
Landau-Ginzburg models, the analogue of the superpotential interactions are
described in superspace in the form
\begin{displaymath}
\sum_{i} \int d \theta  \Lambda^{i} J_{i}(\Phi),
\end{displaymath}
where the $J_{\alpha}$ are a set of holomorphic functions and
$\Lambda^{\alpha}$ are Fermi superfields (forming half of a (2,2) chiral
superfield).  This reduces to a (2,2) superpotential in the special
case that $J_i = \partial_i W$ for some holomorphic function $W$.

Our proposal for the $(0,2)$ Toda-like mirror of the $A/2$ model on
$dP_1 = {\mathbb F}_1$ with a deformation of the tangent bundle is
defined by
\begin{align}
J_1 &=   a X_1 + \mu_{AB} (X_3 - X_1) + b \frac{(X_3 - X_1)^2}{X_1} 
- \frac{q_1}{X_1 (\gamma_1 X_1 + \gamma_2 (X_3 - X_1) )}, \label{eq:dp1-02-1} \\
J_2 &=   a X_1 + \mu_{AB} (X_3 - X_1) + b \frac{(X_3 - X_1)^2}{X_1} 
-  \frac{q_1}{X_1 (\gamma_1 X_1 + \gamma_2 (X_3 - X_1) )}  \nonumber \\
&\quad \quad \quad  + X_3^{-1} \bigg( \Big(\gamma_1 X_1 + \gamma_2 (X_3 - X_1) \Big)
\Big( \alpha_1 X_1 + \alpha_2(X_3 - X_1) \Big) \bigg) - \frac{q_2}{X_3}. \label{eq:dp1-02-2} 
\end{align}
(Because the $J$'s have poles away from origins, we interpret the resulting
action in a low-energy effective field theory sense, as discussed in the
introduction.)  

We have chosen the labels on the $J$'s to match $q$'s, but that also means
they are slightly inconsistent with bosons on the (2,2) locus.  Here,
for example, $J_2$ on the (2,2) locus corresponds to the $Y_3$ derivative
of $W$.

It is straightforward to check that the $J$'s above have the correct
(2,2) locus, and that they are invariant under the $U(1)_V$ but the $U(1)_A$
symmetry is classically broken in the fashion expected.

Previously we gave two forms for the $B$-twisted Landau-Ginzburg mirror to
$dP_1$, on the (2,2) locus.  So far, we have given the mirror that reduces
on the (2,2) locus to the first form.  An expression for a (0,2) mirror
that reduces on the (2,2) locus to the second form is
\begin{align}
J_1 &=  a X_1 + \mu_{AB} X_4 + b \frac{X_4^2}{X_1}  
- \frac{q_1}{q_2} \frac{\alpha_1 X_1 + \alpha_2 X_4}{X_1},  \label{eq:dp1_01-3} \\
J_2 &=  \alpha_2 \gamma_2 X_4 + \alpha_1 \gamma_1 \frac{X_1^2}{X_4}
 + \frac{q_1}{q_2} \frac{ (\alpha_1 X_1 + \alpha_2 X_4) (\gamma_1 \alpha_2
 + \gamma_2 \alpha_1) }{a X_1 + \mu_{AB} X_4 + b X_4^2 X_1^{-1}}
 - \frac{q_2}{X_4}. 
\label{eq:dp1_01-4}
\end{align}
As above, 
we have chosen subscripts on the $J$'s to match $q$'s, which means that
$J_2$ on the (2,2) locus corresponds to the $Y_4$ derivative of $W$.

As above, it is straightforward to 
check that the $J$'s above have the correct
(2,2) locus, and that they are invariant under the $U(1)_V$ but the $U(1)_A$
symmetry is classically broken in the fashion expected.

We will check our proposal by arguing that all genus zero $A/2$ model
correlation functions will match those of the $B/2$-twisted mirror
Landau-Ginzburg theory given above, using a variation of an argument
in \cite{Chen:2016tdd} which can be adapted to apply to potential
(0,2) Landau-Ginzburg model mirrors to any toric variety realized as a
GLSM.

Given a $B/2$ Landau-Ginzburg model with a superpotential $J_i$, the genus zero correlation functions are given by \cite{Melnikov:2007xi}
\begin{equation}
\langle \phi^{i_1}(x_1) \dots \phi^{i_k}(x_k) \rangle = \sum_{J_i (\phi) = 0}  \phi^{i_1}(x_1) \dots \phi^{i_k}(x_k) \left[\det_{i,j} J_{i,j} \right]^{-1}, \label{eq: Bcorr}
\end{equation}
where the sum is taken over the classical vacua.

From \cite{McOrist:2008ji}, the one-loop effective theory is described by the 
following $J$ functions in general:
\begin{displaymath}
{\cal J}_a = \ln \left[ q_a^{-1} \prod_{\alpha} 
Q_{(\alpha)}^{q_{(\alpha)}^a}\right] ,
\end{displaymath}
where $Q_{(\alpha)}$ encodes the tangent bundle deformations (as opposed
to gauge charges).
In the present case of $dP_1$, the superpotential is given by
\begin{align}
\mathcal{J}_1 &= \ln \big[ q_1^{-1} \det (A \psi + B \tilde{\psi}) 
(\psi \gamma_1 + \tilde{\psi} \gamma_2) \big], \\
\mathcal{J}_2 &= \ln \big[ q_2^{-1} (\psi \gamma_1 + \tilde{\psi} \gamma_2) 
(\psi \alpha_1 + \tilde{\psi} \alpha_2) \big],
\end{align}
and the correlation functions are given by
\begin{displaymath}
\langle f(\psi, \tilde{\psi}) \rangle = \sum_{\mathcal{J}=0} f(\psi, \tilde{\psi}) \Bigg[ \det_{a,b} \mathcal{J}_{a,b} \prod_{\alpha} Q_{(\alpha)}   \Bigg]^{-1} .
\end{displaymath}
Comparing to the formula calculating the correlation functions of Toda dual Landau-Ginzburg model~(\ref{eq: Bcorr}), in order to claim the correlation functions match, we only need to verify
\begin{equation}
\det | J_{i,j}| = \det_{a,b} | \mathcal{J}_{a,b} | \prod_{\alpha} Q_{(\alpha)} \label{eq: match}
\end{equation}
on the space of vacua after identifying $X_1$ with $\psi$ and $X_2$ with $\tilde{\psi}$. Expanding the right side of above formula, we get
\[
\det 
\begin{bmatrix}
\frac{1}{Q_{(s)}^n} \partial_{\psi} (Q_{(k)} Q_{(s)}) & \frac{1}{Q_{(s)}^n} \partial_{\tilde{\psi}} (Q_{(k)} Q_{(s)}) \\
\partial_{\psi} ( Q_{(s)} Q_{(t)}) & \partial_{\tilde{\psi}} ( Q_{(s)} Q_{(t)})  \\
\end{bmatrix} .
\]
One can then easily verify equation~(\ref{eq: match}) holds on the space of 
vacua with $X_1 \sim \psi$ and $X_3 \sim n \psi + \tilde{\psi}$ for both
of the presentations of (0,2) mirrors we have given here.

We will use analogous arguments throughout this paper to compare
genus zero correlation functions in proposed (0,2) mirrors to the original
$A/2$ theories, but for brevity in later sections will only mention the
result, not walk through the details of the computation.

So far we have checked that the genus zero 
correlation functions in this proposed
(0,2) mirror to $dP_1$ match those of the original $A/2$-twisted theory.
In the next section, we will check that there is an analogue of a blowdown
in the mirror.  In later sections we will describe proposals for (0,2)
mirrors to
higher del Pezzo surfaces that blow down to this proposal, and we will
also describe a family of proposals for (0,2) mirrors to Hirzebruch
surfaces that include the proposal of this section for
$dP_1 = {\mathbb F}_1$ as well as our earlier proposal for ${\mathbb P}^1
\times {\mathbb P}^1$ as special cases.

\subsubsection{Consistency check:  mirrors of blowdowns}

\begin{figure}[htb]
\begin{center}
\begin{picture}(150,150)
\LongArrow(75,75)(150,75)  \Text(140,80)[b]{$(1,0)$}
\LongArrow(75,75)(75,150)  \Text(80,140)[l]{$(0,1)$}
\LongArrow(75,75)(0,0)      \Text(10,30)[b]{$(-1,-1)$}
\DashLine(75,75)(75,0){2}
\Vertex(75,75){2}
\end{picture}
\end{center}
\caption{A toric fan of $\mathbb{P}^2$ can be 
obtained by removing the edge $(0,-1)$ from the toric fan of $dP_1$.
\label{fig:dp1-p2}}
\end{figure}
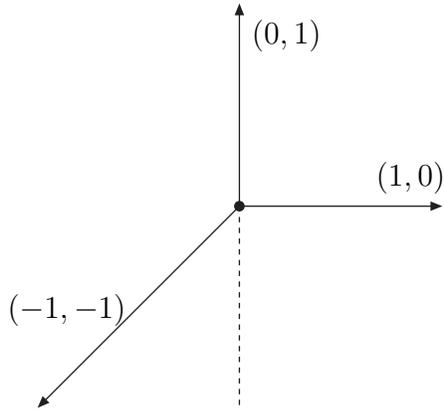

Geometrically, $dP_1$ can be blown down to ${\mathbb P}^2$, which is visible
in the toric fan in figure~\ref{fig:dp1-p2}
by removing the edge $(0,-1)$.  In the GLSM, although in general
K\"ahler moduli of non-Calabi-Yau manifolds need not correspond to operators
in the physical theory, it is nevertheless straightforward to see that there
is an analogous limit\footnote{
The same statement will be true of the other blowdown examples considered
in this paper -- all involving blowups of Fano spaces at smooth points.
} in which one recovers ${\mathbb P}^2$.
The (2,2) mirror of this
blowdown is manifest that we only need to take the
limit $q_2 \to 0$ in (\ref{eq:22_sp_one}), which reduces to the Toda dual 
superpotential of $\mathbb{P}^2$. 

In this section we will show that the blowdown limit of the (0,2) mirror
of $dP_1$ with a tangent bundle deformations is also 
equivalent (as a UV theory) to the mirror of $\mathbb{P}^2$.
This will provide a consistency test of our proposed (0,2) mirror.

To that end, it will be helpful to first revisit the (2,2) case, albeit in 
(0,2) language.
Recall
\begin{equation*}
W = \Lambda^1 J_1 + \Lambda^2 J_2,
\end{equation*}
where $\Lambda^i, i =1,2$ are Fermi superfields, and
\begin{align*}
J_1 &= X_1 - \frac{q_1}{X_1 X_3}, \\
J_2 & = X_3 - \frac{q_1}{X_1 X_3}.
\end{align*}
We can rewrite the (0,2) superpotential as follows,
\begin{equation*}
W = \tilde{\Lambda}^1 \tilde{J}_1 + \tilde{\Lambda}^2 \tilde{J}_2,
\end{equation*}
where
\begin{equation*}
 \tilde{\Lambda}^1 =  \Lambda^1 + \Lambda^2, 
\quad  \tilde{\Lambda}^2 = \Lambda^2,
 \end{equation*}
 and
\begin{align*}
  \tilde{J}_1 &= J_1 = X_1 - \frac{q_1}{X_1 X_3}, \\
  \tilde{J}_2 &= J_2 - J_1 = X_3 - X_1.
\end{align*}
Then, one can integrate out the Fermi superfield $\tilde{\Lambda}^2$ and obtain a constraint,
\begin{equation*}
X_1  = X_3.
\end{equation*}
Plugging the constraint back in, we get 
\begin{equation}
  W = \tilde{\Lambda}^1 \tilde{J}_1 
= \tilde{\Lambda}^1\left( X_1 - \frac{q_1}{X_1^2} \right). \label{eq:p2}\\
\end{equation}

Now let us analyze the (0,2) superpotential of $dP_1$ in the 
blowdown limit $q_2 \to 0$,
\begin{equation*}
W = \Lambda^1 J_1 + \Lambda^2 J_2,
\end{equation*}
where $\Lambda^i, i =1,2$ are Fermi superfields, and 
\begin{align*}
J_1 &=   a X_1 + \mu_{AB} (X_3 - X_1) + b \frac{(X_3 - X_1)^2}{X_1} 
- \frac{q_1}{X_1 (\gamma_1 X_1 + \gamma_2 (X_3 - X_1) )}, \\
J_2 &=   a X_1 + \mu_{AB} (X_3 - X_1) + b \frac{(X_3 - X_1)^2}{X_1} 
-  \frac{q_1}{X_1 (\gamma_1 X_1 + \gamma_2 (X_3 - X_1) )} \\
&\quad \quad \quad  + X_3^{-1} \bigg( \Big(\gamma_1 X_1 + \gamma_2 (X_3 - X_1) \Big)
\Big( \alpha_1 X_1 + \alpha_2(X_3 - X_1) \Big) \bigg). 
\end{align*}

We can rewrite it as
\begin{equation*}
W = \tilde{\Lambda}^1 \tilde{J}_1 + \tilde{\Lambda}^2 \tilde{J}_2,
\end{equation*}
where
\begin{equation*}
 \tilde{\Lambda}^1 =  \Lambda^1 + \Lambda^2, 
\quad  \tilde{\Lambda}^2 = \Lambda^2,
 \end{equation*}
 and
 \begin{align*}
  \tilde{J}_1 &= J_1 =  a X_1 + \mu_{AB} (X_3 - X_1) 
+ b \frac{(X_3 - X_1)^2}{X_1} 
- \frac{q_1}{X_1 (\gamma_1 X_1 + \gamma_2 (X_3 - X_1) )}, \\
  \tilde{J}_2 &= J_2 - J_1 = 
+ X_3^{-1} \left( \left(\gamma_1 X_1 + \gamma_2 (X_3 - X_1) \right) 
\left( \alpha_1 X_1 + \alpha_2(X_3 - X_1) \right) \right).
  \end{align*}

Then, we integrate out $\tilde{\Lambda}^2$ and obtain the following 
constraint on $X_1, X_3$:
\begin{equation*}
\tilde{J}_2 =  X_3^{-1} 
\left( \left(\gamma_1 X_1 + \gamma_2 (X_3 - X_1) \right) 
\left( \alpha_1 X_1 + \alpha_2(X_3 - X_1) \right) \right) = 0.
\end{equation*}
Notice that $\gamma_1 X_1 + \gamma_2 (X_3 - X_1) \neq 0$ 
since it is in the denominator of $J_1$. Solving the constraint, 
one obtain the relation 
\begin{displaymath}
X_3 = \frac{\alpha_2 - \alpha_1}{\alpha_2} X_1,
\end{displaymath}
where for simplicity we have assumed $\alpha_2 \neq 0$.

Lastly, plugging the above relation back into $\tilde{J}_1$, we find
\begin{equation*}
W = \tilde{\Lambda}^1 \tilde{J}_1 
= \tilde{\Lambda}^1 \left(  \left( a - \mu_{AB} \alpha_1 \alpha_2^{-1} + b  \alpha_1^2 \alpha_2^{-2} \right) X_1 
- \frac{q_1}{\left( \gamma_1 - \gamma_2 \alpha_1 \alpha_2^{-1}\right) X_1^2} \right).
\end{equation*}
(We assume for simplicity that $\gamma_1 \neq \gamma_2 \alpha_1/\alpha_2$.)
One can easily see the above superpotential is equivalent to (\ref{eq:p2})
for the mirror to ${\mathbb P}^2$, after suitable field redefinitions.
Thus, as expected, mirrors and blowdowns commute with one another.

\subsection{The second del Pezzo surface, $dP_2$}

\subsubsection{Review of the (2,2) mirror}  \label{sect:dp2:rev22}

The next del Pezzo surface, $dP_2$, is $\mathbb{P}^2$ blown up at two points, 
which can be described as a toric variety by a fan with edges 
$(1,0)$, $(0,1)$, $(-1,-1)$, $(0,-1)$, $(-1,0)$. The gauged linear sigma model 
has five chiral superfields $\phi_i, i=1 \dots 5$ which are 
charged under the gauge group $U(1)^3$ as follows:
\[
\begin{tabular}{ c c c c c}
(1,0) & (-1,-1) & (0,1) & (0,-1) & (-1,0) \\
\hline
1 & 1 & 1 & 0 & 0 \\
0 & 0 & 1 & 1 & 0 \\
1 & 0 & 0 & 0 & 1 
\end{tabular}
\]
The quantum cohomology relations of the $A$-twisted theory are
\begin{align*}
(\psi_1 + \psi_3) \psi_1 (\psi_1 + \psi_2) &= q_1, \\
(\psi_1 + \psi_2) \psi_2 &= q_2, \\
(\psi_1 + \psi_3) \psi_3 &= q_3.
\end{align*}

As reviewed in section~\ref{sect:review:22},
the superpotential of the (2,2) mirror theory is 
\begin{displaymath}
W = \sum_{i=1}^5 \exp ( + Y_i)
\end{displaymath}
where the $Y_i$ obey constraints
\begin{displaymath}
Y_1 + Y_2 + Y_3 = r_1, \quad Y_3 + Y_4 = r_2, \quad Y_1 + Y_5 = r_3.
\end{displaymath}
One solution is to solve the constraints for $Y_4$ and $Y_5$.  Defining
$X_4 = \exp ( Y_4)$ and $X_5 = \exp ( Y_5)$,
the superpotential is then
\begin{equation}
W = X_4 + X_5 + \frac{q_3}{X_5} + \frac{q_1}{q_2 q_3} X_4 X_5 + \frac{q_2}{X_4}. \label{eq: 22_sp_two}
\end{equation}
(The mirror map relates  
$\psi_2 \sim X_4$ and $X_5 \sim \psi_3$.)

However, we will not use the form of the Toda dual above when building
(0,2) deformations.  Instead, we will use
an alternative form of the Toda dual, which is obtained 
by retaining an explicit Lagrange multiplier $Z$, 
so that one of the constraints naturally embeds into the superpotential,
\begin{equation}
W = X_1 + X_3 + X_5 + \frac{q_1}{X_1 X_3} + \frac{q_2}{X_3}
 + Z \left( 1 - \frac{q_3}{X_1 X_5} \right),  \label{eq: alt_22_sp-1}
\end{equation}
for $X_i = \exp(Y_i)$.
(If we solve the constraint by taking $X_1 = q_3/X_5$, then this form
can be related to the previous expression by the holomorphic
coordinate transformation $X_4 = q_2/X_3$.)
On the space of vacua which is given by,
\begin{align*}
 X_1 \partial_1 W &=  X_1 - \frac{q_1}{X_1 X_3} + Z \frac{q_3}{X_1 X_5} = 0, \\
 X_3 \partial_3 W &=  X_3 - \frac{q_1}{X_1 X_3} - \frac{q_2}{X_3} = 0, \\
 X_5 \partial_5 W &=  X_5 + Z \frac{q_3}{X_1 X_5} = 0, \\
\partial_Z W &= 1 - \frac{q_3}{X_1 X_5} = 0,
\end{align*}
where
\begin{displaymath}
\partial_i = \frac{\partial}{\partial X_i}.
\end{displaymath}
The quantum cohomology relations are satisfied with the identifications
\begin{displaymath}
X_1 \sim \psi_1 + \psi_3, \quad X_3 \sim \psi_1 + \psi_2, \quad X_5 \sim \psi_3,
\end{displaymath}
It is straightforward to check that the vector R symmetry is preserved,
but the axial R symmetry is broken, as expected.
Furthermore, it is easy to check that all correlation functions of the 
alternative description match\footnote{
Technically, keeping an explicit Lagrange multiplier turns out to
introduce a sign in correlation functions, which can easily be accounted for.
} those of $A$-twisted theory for $dP_2$.
This alternative description turns out to be convenient for
constructing ansatzes for
(0,2) mirrors.

The choice of constraint embedding in the superpotential should be arbitrary. 
For example, we could also take the mirror superpotential to be
\begin{equation} 
W = X_1 + X_2 + X_3 + \frac{q_2}{X_3} + \frac{q_3}{X_1} + Z \left( 1 - \frac{q_1}{X_1 X_2 X_3} \right),  \label{eq: alt_22_sp}
\end{equation}
with vacua, 
\begin{align*}
 X_1 \partial_1 W &=  X_1 - \frac{q_3}{X_1} + Z \frac{q_1}{X_1 X_2 X_3} = 0, \\
 X_2 \partial_2 W &=  X_2 + Z \frac{q_1}{X_1 X_2 X_3} = 0, \\ 
 X_3 \partial_3 W &=  X_3 - \frac{q_2}{X_3} +  Z \frac{q_1}{X_1 X_2 X_3} = 0, \\
\partial_Z W & = 1 - \frac{q_1}{X_1 X_2 X_3} = 0.
\end{align*}
This can be related to the previous form using the holomorphic
coordinate transformation $X_2 = (q_1/q_3) (X_5/X_3)$.
We will also describe (0,2) deformations of this presentation.

\subsubsection{(0,2) deformations and proposed (0,2) mirrors} \label{sec: dp2-02}

The $(0,2)$ deformation of $dP_2$ is defined by fifteen complex numbers $\alpha_i$, $\beta_j$, $\gamma_k$, $\delta_m$, $\epsilon_n$, with $i,j,k,m,n = 1,2,3$, which define a deformation ${\cal E}$ of the tangent bundle as follows:
\begin{equation*}
\begin{split}
0 \: \longrightarrow {\cal O}^3 \: \xrightarrow{E} \: {\cal O}(1,0,1) \oplus {\cal O}(1,0,0) \oplus {\cal O}(1,1,0) \oplus  {\cal O}(0,1,0) \oplus {\cal O}(0,0,1) \: \\
\quad  \longrightarrow \: {\cal E} \: \longrightarrow 0 ,
\end{split}
\end{equation*}
where
\begin{displaymath}
E = 
\begin{bmatrix}
\alpha_1 s_1 & \alpha_2 s_1 & \alpha_3 s_1 \\
\beta_1 s_2 & \beta_2 s_2 & \beta_3 s_2 \\
\gamma_1 s_3 & \gamma_2 s_3 & \gamma_3 s_3 \\
\delta_1 s_4 & \delta_2 s_4 & \delta_3 s_4 \\
\epsilon_1 s_5 & \epsilon_2 s_5 & \epsilon_3 s_5 
\end{bmatrix}.
\end{displaymath}
${\cal E}$ reduces to the tangent bundle when
\begin{align*}
 &\alpha_1 = 1, \quad \alpha_2 = 0, \quad \alpha_3 = 1, \\
 &\beta_1 =1, \quad \beta_2 = \beta_3 = 0, \\
 &\gamma_1 = \gamma_2 =1, \quad \gamma_3 = 0, \\
 &\delta_1 = 0, \quad \delta_2 = 1, \quad \delta_3 = 0, \\
 &\epsilon_1 = \epsilon_2 = 0, \quad \epsilon_3 = 1.
\end{align*}

The quantum sheaf cohomology relations are
\begin{align}
 Q_{(1)} Q_{(2)} Q_{(3)} & = q_1,  \label{eq: 2blowup_qsc1} \\
 Q_{(3)} Q_{(4)} & = q_2, \label{eq: 2blowup_qsc2}\\
 Q_{(1)} Q_{(5)} & = q_3, \label{eq: 2blowup_qsc3}
\end{align}
where
\begin{align*}
Q_{(1)} &= \sum_{i=1}^3 \alpha_i \psi_i, \quad Q_{(2)} = \sum_{i=1}^3 \beta_i \psi_i, \quad Q_{(3)} = \sum_{i=1}^3 \gamma_i \psi_i,  \\
& \qquad Q_{(4)} = \sum_{i=1}^3 \delta_i \psi_i, 
\quad Q_{(5)} = \sum_{i=1}^3 \epsilon_i \psi_i.
\end{align*}

We will propose below two (0,2) Toda-like mirrors based on the
(2,2) mirrors (\ref{eq: alt_22_sp-1}) and (\ref{eq: alt_22_sp})
which have a Lagrange multiplier (labelled $Z$).

Our first (0,2) mirror proposal for $dP_2$ is defined by the following
four holomorphic functions
\begin{align}
J_1 &= - \frac{q_1}{X_1 (\gamma \cdot X)} + Z \frac{q_3}{X_1 (\epsilon \cdot X)}
 + \frac{(\alpha \cdot X) (\epsilon \cdot X)}{X_1} 
+ \frac{ (\alpha \cdot X) (\beta \cdot X) }{X_1}, \label{eq:dp2-alt-1-1}\\
J_3 &= - \frac{q_2}{X_3} - \frac{q_1}{X_1 (\gamma \cdot X)} 
+ \frac{(\alpha \cdot X) (\beta \cdot X) }{X_1} 
+ \frac{ (\gamma \cdot X) (\delta \cdot X) }{X_3}, \\
J_5 &=  (\epsilon \cdot X) + Z \frac{q_3}{ (\alpha \cdot X) (\epsilon \cdot X)}, \\
J_Z &= \frac{ ( \epsilon \cdot X ) }{X_5} - \frac{q_3}{ X_5 ( \alpha \cdot X ) }, \label{eq:dp2-alt-1-4}
\end{align}
where on the (2,2) locus, it can be shown that
\begin{displaymath}
J_1 \: = \: \frac{\partial W}{\partial Y_1}, \: \: \:
J_3 \: = \: \frac{\partial W}{\partial Y_3}, \: \: \:
J_5 \: = \: \frac{\partial W}{\partial Y_5}, \: \: \:
J_Z \: = \: \frac{\partial W}{\partial Z},
\end{displaymath}
for $W$ defined by (\ref{eq: alt_22_sp-1}),
and
\begin{align*}
( \alpha \cdot X ) & = \alpha_1 (X_1 - X_5) + \alpha_2 (X_3 - X_1 + X_5) + \alpha_3 X_5, \\
& \; \; \vdots \\
( \epsilon \cdot X ) & = \epsilon_1 (X_1 - X_5) + \epsilon_2 (X_3 - X_1 + X_5) + \epsilon_3 X_5.
\end{align*}
Because the $J$'s have poles away from origins, we interpret the resulting
action in a low-energy effective field theory sense, as discussed in the
introduction.  It is straightforward to check that this proposal has the
correct (2,2) locus.

Our second (0,2) mirror ansatz for $dP_2$ is defined by the following
data:
\begin{align}
J_1&= -\frac{q_3}{X_1} + \frac{( \alpha \cdot X)( \epsilon \cdot X)}{X_1}
 + ( \beta \cdot X) + Z \, \frac{q_1}{( \alpha \cdot X)( \beta \cdot X)( \gamma \cdot X)}, \label{eq:dp2_alt_1} \\
J_2&=  (\beta \cdot X) + Z \, \frac{q_1}{( \alpha \cdot X)( \beta \cdot X)( \gamma \cdot X)}, \\
J_3 &= - \frac{q_2}{X_3} 
+ \frac{( \gamma \cdot X)( \delta \cdot X)}{X_3} + ( \beta \cdot X) 
+ Z \, \frac{q_1}{( \alpha \cdot X)( \beta \cdot X)( \gamma \cdot X)}, \\
J_Z &=  + \frac{( \beta \cdot X)}{X_2} 
- \frac{q_1}{X_2 ( \alpha \cdot X) ( \gamma \cdot X)}, \label{eq:dp2_alt_4}
\end{align}
where
\begin{align*}
( \alpha \cdot X) &= \alpha_1 X_2 + \alpha_2 ( X_3 - X_2) + \alpha_3 ( X_1 - X_2), \\
&\quad \vdots \\
( \epsilon \cdot X) &= \epsilon_1 X_2 + \epsilon_2 ( X_3 - X_2) + \epsilon_3 ( X_1 - X_2).
\end{align*}
On the $(2,2)$ locus, the above data reduces to (\ref{eq: alt_22_sp})
(in the sense that each $J_i$ becomes a suitable derivative of $W$).

With the identifications,
\[
X_1 \sim \psi_1+\psi_3, \quad X_2 \sim \psi_1, \quad X_3 \sim \psi_1+\psi_2, 
\quad X_5 \sim \psi_3,
 \] 
both proposals pass our standard consistency checks: 
the quantum sheaf cohomology relations are satisfied on the vacua,
the $U(1)_V$ symmetry is unbroken but the $U(1)_A$ broken classically, 
and all correlation functions match those of $A/2$ twisted theory as before.

\subsubsection{Consistency check: mirrors of blowdowns to $dP_1$}

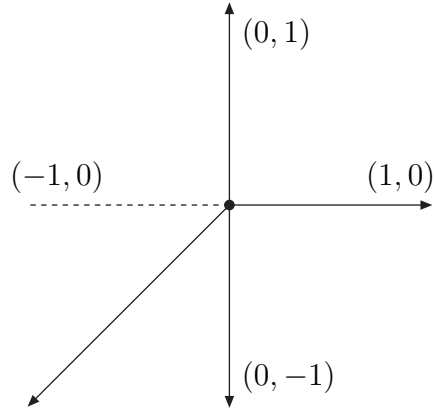
\begin{figure}[htb]
\begin{center}
\begin{picture}(150,150)
\LongArrow(75,75)(150,75)  \Text(140,80)[b]{$(1,0)$}
\LongArrow(75,75)(75,150)  \Text(80,140)[l]{$(0,1)$}
\LongArrow(75,75)(0,0)
\LongArrow(75,75)(75,0)  \Text(80,10)[l]{$(0,-1)$}
\DashLine(75,75)(0,75){2}  \Text(10,80)[b]{$(-1,0)$}
\Vertex(75,75){2}
\end{picture}
\end{center}
\caption{A toric fan for $dP_1$ can be obtained by removing 
the edge $(-1,0)$ from the toric fan for $dP_2$. \label{fig:dp2-dp1}}
\end{figure}

The del Pezzo surface $dP_2$ can be blown down to $dP_1$, 
which one can see from the toric fan by removing the edge $(-1,0)$
in figure~\ref{fig:dp2-dp1}.  (Moreover, essentially because we are discussing
blowups of smooth points on Fano varieties, the UV phases of the GLSMs
are the geometries described here, so in the cases described
here there are no subtlties involving the GLSM giving results for
unexpected geometries.)

On the (2,2) locus, the mirror of the blowdown from $dP_2$ to $dP_1$
is described in the Toda dual theory (\ref{eq: alt_22_sp}) by taking the limit 
$q_3 \to 0$ after integrating out the Lagrange multiplier $Z$. 
Next, we will analyze both of the proposed (0,2) mirror theories in section (\ref{sec: dp2-02}) under the same blowdown limit.

First, to illustrate the method,
let us explain how to explicitly follow 
the blowdown in the (2,2) Toda dual
(\ref{eq: alt_22_sp}).
That (2,2) superpotential can be rewritten in the $(0,2)$ language as follows,
\begin{equation*}
\int \; \mathrm{d} \theta \; W(\Phi) =  \sum_i \int \; \mathrm{d} \theta \; \Lambda^i J_i (\Phi),
\end{equation*}
where the $\Lambda^i$ are Fermi superfields and the $J_i$ are derivatives
of $W$, which in the current case are given by 
\begin{align*}
J_1 &=  X_1 - \frac{q_3}{X_1} + Z \frac{q_1}{X_1 X_2 X_3} , \\
J_2 &=  X_2 + Z \frac{q_1}{X_1 X_2 X_3} , \\ 
J_3 &=  X_3 - \frac{q_2}{X_3} +  Z \frac{q_1}{X_1 X_2 X_3} , \\
J_Z & = 1 - \frac{q_1}{X_1 X_2 X_3} .
\end{align*}
To integrate out the Lagrange multiplier $Z$, 
one integrates out the Fermi field $\Lambda^Z$ corresponding
to $J_Z = \partial_Z W$, which implies 
\begin{equation*}
J_Z = 1 - \frac{q_1}{X_1 X_2 X_3} = 0,
\end{equation*}
or
\begin{equation*}
X_2 = \frac{q_1}{X_1 X_3}. 
\end{equation*}
As one might expect, the above constraint is the same constraint arising from
integrating out the Lagrange multiplier $Z$ in (\ref{eq: alt_22_sp}). 
Imposing this constraint, the remaining $J$ functions become
\begin{align*}
J_1 &= X_1 - \frac{q_3}{X_1} + Z, \\
J_2 &=  \frac{q_1}{X_1 X_3} + Z, \\
J_3 &= X_3 + Z - \frac{q_2}{X_3}. \\
\end{align*}
Since we have removed explicit $X_2$ dependence from the $J_i$ above,
we should also integrate out the Fermi field $\Lambda^2$ corresponding 
to $J_2 = - X_2 \partial_{X_2} W$, which implies
\begin{equation*}
Z = - \frac{q_1}{X_1X_3}.
\end{equation*}
Applying the constraint above, one reaches the form
\begin{align*}
J_1 &= X_1 - \frac{q_3}{X_1} - \frac{q_1}{X_1 X_3}, \\
J_3 &= X_3 - \frac{q_1}{X_1 X_3} - \frac{q_2}{X_3}.
\end{align*}
Finally, taking the limit $q_3 \rightarrow 0$, we see that the $J$ functions
above precisely coincide with those for the (2,2) Toda dual of 
$dP_1$ presented in (\ref{eq:22_sp_one}).

Now that we have illustrated the method, 
let us analyze the mirror of the 
$(0,2)$ theory (\ref{eq:dp2-alt-1-1})-(\ref{eq:dp2-alt-1-4}) 
in the blowdown limit $q_3 \to 0$. 
After integrating out the Lagrange multiplier, one obtains the constraints
\begin{align*}
J_5 &=  (\epsilon \cdot X) + Z \frac{q_3}{ (\alpha \cdot X) (\epsilon \cdot X)} = 0, \\
J_Z &= \frac{ ( \epsilon \cdot X ) }{X_5} - \frac{q_3}{ X_5 ( \alpha \cdot X ) } = 0, 
\end{align*}
where
\begin{align*}
( \alpha \cdot X ) & = \alpha_1 (X_1 - X_5) + \alpha_2 (X_3 - X_1 + X_5) + \alpha_3 X_5, \\
& \; \; \vdots \\
( \epsilon \cdot X ) & = \epsilon_1 (X_1 - X_5) + \epsilon_2 (X_3 - X_1 + X_5) + \epsilon_3 X_5.
\end{align*}
In the limit $q_3 \to 0$, the constraint $J_Z=0$ implies
\begin{equation*}
X_5 = \frac{\epsilon_1 X_1 + \epsilon_2 (X_3 - X_1)}{\epsilon_1 - \epsilon_2 - \epsilon_3}.
\end{equation*}
(For simplicity, we assume $\epsilon_1 - \epsilon_2 - \epsilon_3 \neq 0$.)
The mirror blowdown is then given by,
\begin{align*}
J_1 &= - \frac{q_1}{X_1 (\Gamma_1 X_1 + \Gamma_2 (X_3 - X_1) ) } 
\\
& \qquad 
+ \frac{ (A_1 X_1 + A_2 (X_3 - X_1) ) (B_1 X_1 + B_2 (X_3 - X_1) )}{X_1}, \\
J_3 &= - \frac{q_1}{X_1 (\Gamma_1 X_1 + \Gamma_2 (X_3 - X_1) ) } 
\\
& \qquad
+ \frac{ (A_1 X_1 + A_2 (X_3 - X_1) ) (B_1 X_1 + B_2 (X_3 - X_1) )}{X_1} \\
& \qquad  + \frac{ (\Gamma_1 X_1 + \Gamma_2 (X_3 - X_1) ) (\Delta_1 X_1 + \Delta_2 (X_3 - X_1) )}{X_3} -  \frac{q_2}{X_3}, 
\end{align*}
where,
\begin{align*}
A_1 & = \frac{ \epsilon_1 ( \alpha_2 + \alpha_3) - \alpha_1 (\epsilon_2 + \epsilon_3)}{\epsilon_1 - \epsilon_2 - \epsilon_3}, 
\quad
A_2 = \frac{ \epsilon_2 ( \alpha_3 - \alpha_1) + \alpha_2 ( \epsilon_1 - \epsilon_3) }{\epsilon_1 - \epsilon_2 - \epsilon_3}, \\
B_1 & = \frac{ \epsilon_1 ( \beta_2 + \beta_3) - \beta_1 (\epsilon_2 + \epsilon_3) }{\epsilon_1 - \epsilon_2 - \epsilon_3}, 
\quad
B_2 = \frac{ \epsilon_2 ( \beta_3 - \beta_1) + \beta_2 ( \epsilon_1 - \epsilon_3) }{\epsilon_1 - \epsilon_2 - \epsilon_3}, \\
\Gamma_1 & = \frac{ \epsilon_1 ( \gamma_2 + \gamma_3) - \gamma_1 (\epsilon_2 + \epsilon_3) }{\epsilon_1 - \epsilon_2 - \epsilon_3}, 
\quad
\Gamma_2 = \frac{ \epsilon_2 ( \gamma_3 - \gamma_1) + \gamma_2 ( \epsilon_1 - \epsilon_3) }{\epsilon_1 - \epsilon_2 - \epsilon_3}, \\
\Delta_1 & = \frac{ \epsilon_1 ( \delta_2 + \delta_3) - \delta_1 (\epsilon_2 + \epsilon_3) }{\epsilon_1 - \epsilon_2 - \epsilon_3}, 
\quad
\Delta_2 = \frac{ \epsilon_2 ( \delta_3 - \delta_1) + \delta_2 ( \epsilon_1 - \epsilon_3) }{\epsilon_1 - \epsilon_2 - \epsilon_3}.
\end{align*}
One can see that the resulting superpotential is the same as the superpotential (\ref{eq:dp1-02-1}), (\ref{eq:dp1-02-2}) after adjusting the parameters as follows,
\begin{align*}
&a = A_1 B_1, \quad b = A_2 B_2, \quad \mu_{AB} = A_1 B_2 + A_2 B_1, \\
&\gamma_1 = \Gamma_1, \quad  \gamma_2 = \Gamma_2, \\
&\alpha_1 = \Delta_1, \quad \alpha_2 = \Delta_2.
\end{align*}
Thus, as expected, the (0,2) mirror to the blowdown, is the blowdown limit
of the mirror.  This provides a consistency check on the form of the
proposed mirror.

Next, we repeat the analysis for 
the second form of
the $(0,2)$ mirror (\ref{eq:dp2_alt_1})-(\ref{eq:dp2_alt_4}).
We first obtain the constraints
\begin{align*}
J_2&=  (\beta \cdot X) + Z \, \frac{q_1}{( \alpha \cdot X)( \beta \cdot X)( \gamma \cdot X)} = 0, \\
J_Z &=   \frac{( \beta \cdot X)}{X_2} - \frac{q_1}{X_2 ( \alpha \cdot X) ( \gamma \cdot X)} = 0.
\end{align*}
In principle, one can use these constraints to eliminate the 
dependence on $X_2$ and $Z$ in the remaining $J$ functions. 
Then, taking the limit $q_3 \to 0$ one should recover 
the $J$ functions of the $(0,2)$ mirror of $dP_1$. 
However, $J_Z$ is effectively a cubic polynomial in $X_2$, so directly
solving for $X_2$ in arbitrary (0,2) deformations is rather complex.
For simplicity, we will only consider the blowdown in the second form
of the (0,2) mirror for a special
family of deformations, of the form
\begin{displaymath}
\alpha_1 = 1, \quad \alpha_2 = 0, \quad \alpha_3 =1, \quad \gamma_1 =1, 
\quad \gamma_2 =1, \quad \gamma_3 =0,
\end{displaymath}
leaving other deformation parameters arbitrary.

Now, for this family of deformations, the constraints become
\begin{align*}
X_2 &= (\beta_1 - \beta_2 -\beta_3)^{-1} \left( \frac{q_1}{X_1 X_3} - \beta_3 X_1 - \beta_2 X_3 \right) , \\
Z &= - \frac{q_1}{X_1 X_3} = - \beta \cdot X. 
\end{align*}
Plugging back into the other $J$ functions, we find
\begin{align*}
J_1' &= E \; J_1, \\
&=  - \frac{q_1}{X_1 X_3} 
- \frac{\beta_1 \epsilon_3 - \beta_2 \epsilon_3 - \beta_3 \epsilon_1 + \beta_3 \epsilon_2}{\epsilon_1 - \epsilon_2 - \epsilon_3} X_1 
- \frac{\beta_1 \epsilon_2 - \beta_3 \epsilon_2 - \beta_2 \epsilon_1 + \beta_2 \epsilon_3}{\epsilon_1 - \epsilon_2 - \epsilon_3} X_3, \\
J_3' &= \Delta \; J_3, \\
&= - \frac{q_2'}{X_3} - \frac{q_1}{X_1 X_3} 
- \frac{\beta_1 \delta_3 - \beta_2 \delta_3 - \beta_3 \delta_1 + \beta_3 \delta_2}{\delta_1 - \delta_2 - \delta_3} X_1 
\\
& \qquad
- \frac{\beta_1 \delta_2 - \beta_3 \delta_2 - \beta_2 \delta_1 + \beta_2 \delta_3}{\delta_1 - \delta_2 - \delta_3} X_3,
\end{align*}
where
\begin{align*}
E &= - \frac{\beta_1 - \beta_2 - \beta_3}{\epsilon_1 - \epsilon_2 - \epsilon_3} , \\
\Delta &= - \frac{\beta_1 - \beta_2 - \beta_3}{\delta_1 - \delta_2 - \delta_3}, \\
q_2' &= - \frac{\beta_1 - \beta_2 - \beta_3}{\delta_1 - \delta_2 - \delta_3} q_2.
\end{align*}
We assume that
\begin{displaymath}
\delta_1 - \delta_2 - \delta_3 \neq 0, \: \: \:
\epsilon_1 - \epsilon_2 - \epsilon_3 \neq 0.
\end{displaymath}

Note that we rescaled $J_1$ and $J_3$:
the rescaling parameters $E$ and $\Delta$ can always be absorbed in the 
corresponding Fermi fields. We also rescaled $q_2$ to match 
the form of the $J$ functions of $dP_1$. As a result, one can see 
that the $J$ functions reduce to those of 
$dP_1$ in equations~(\ref{eq:dp1-02-1})-(\ref{eq:dp1-02-2}), 
with the parameters related as follows:
\begin{align*}
\gamma_1 &= \gamma_2 = 1, \quad b = 0, \\
a &= - \frac{\beta_1 \epsilon_2 - \beta_2 \epsilon_1 + \beta_1 \epsilon_3 - \beta_3 \epsilon_1}{\epsilon_1 - \epsilon_2 - \epsilon_3}, \\
\mu_{AB} &= - \frac{\beta_1 \epsilon_2 - \beta_3 \epsilon_2 - \beta_2 \epsilon_1 + \beta_2 \epsilon_3}{\epsilon_1 - \epsilon_2 - \epsilon_3},\\
\alpha_1 &= \frac{\beta_1 \epsilon_3 - \beta_3 \epsilon_1 + \beta_1 \epsilon_2 - \beta_2 \epsilon_1}{\epsilon_1 - \epsilon_2 - \epsilon_3}
- \frac{\beta_1 \delta_2 - \beta_3 \delta_1 - \beta_2 \delta_1 + \beta_1 \delta_3}{\delta_1 - \delta_2 - \delta_3}, \\
\alpha_2 &= \frac{\beta_1 \epsilon_2 - \beta_3 \epsilon_2 - \beta_2 \epsilon_1 + \beta_2 \epsilon_3}{\epsilon_1 - \epsilon_2 - \epsilon_3} - \frac{\beta_1 \delta_2 - \beta_3 \delta_2 - \beta_2 \delta_1 + \beta_2 \delta_3}{\delta_1 - \delta_2 - \delta_3}.
\end{align*}

\subsubsection{Consistency check: mirrors of blowdowns to ${\mathbb P}^1 \times {\mathbb P}^1$}

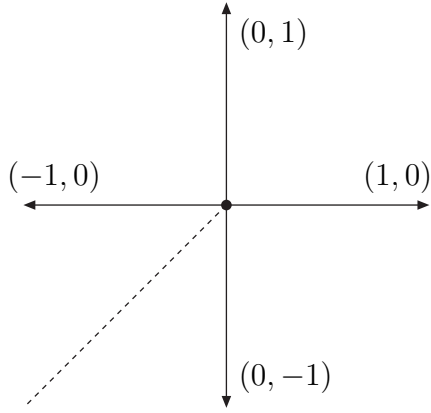
\begin{figure}[htb]
\begin{center}
\begin{picture}(150,150)
\LongArrow(75,75)(150,75)  \Text(140,80)[b]{$(1,0)$}
\LongArrow(75,75)(75,150)  \Text(80,140)[l]{$(0,1)$}
\DashLine(75,75)(0,0){2}
\LongArrow(75,75)(75,0)  \Text(80,10)[l]{$(0,-1)$}
\LongArrow(75,75)(0,75)  \Text(10,80)[b]{$(-1,0)$}
\Vertex(75,75){2}
\end{picture}
\end{center}
\caption{A toric fan for ${\mathbb P}^1 \times {\mathbb P}^1$ can be
obtained by removing the edge $(-1,-1)$ from the toric fan for $dP_2$.
\label{fig:dp2-p1p1} }
\end{figure}

We can also blowdown $dP_2$ to ${\mathbb P}^1 \times {\mathbb P}^1$,
which can be represented in the toric fan we have used previously by
removing the edge $(-1,-1)$, as shown in figure~\ref{fig:dp2-p1p1}.
(As before, since we are discussing Fano varieties, the geometries described
all correspond to UV phases of the GLSMs.)

On the $(2,2)$ locus, the mirror of the blowdown from 
$dP_2$ to ${\mathbb P}^1 \times {\mathbb P}^1$ is described in the 
mirror theory (\ref{eq: alt_22_sp}) by taking the limit $q_1 \to 0$ after 
integrating out the Lagrange multiplier $Z$. Off the $(2,2)$ locus, 
we can follow the same procedure as before, integrating out the 
Lagrange multiplier in (\ref{eq:dp2_alt_1})-(\ref{eq:dp2_alt_4}) and 
taking the limit $q_1 \to 0$ to blow down the $(0,2)$ mirror dual 
$J$ functions of $dP_2$ to $\mathbb{P}^1 \times \mathbb{P}^1$. Integrating out 
the Lagrange multiplier, we obtain the constraints
\begin{align*}
J_2&=  (\beta \cdot X) + Z \, \frac{q_1}{( \alpha \cdot X)( \beta \cdot X)( \gamma \cdot X)} = 0, \\
J_Z &=  \frac{( \beta \cdot X)}{X_2} - \frac{q_1}{X_2 ( \alpha \cdot X) ( \gamma \cdot X)} = 0,
\end{align*}
where
\begin{align*}
( \alpha \cdot X) &= \alpha_1 X_2 + \alpha_2 ( X_3 - X_2) + \alpha_3 ( X_1 - X_2), \\
&\quad \vdots \\
( \epsilon \cdot X) &= \epsilon_1 X_2 + \epsilon_2 ( X_3 - X_2) + \epsilon_3 ( X_1 - X_2).
\end{align*}
In the limit $q_1 \to 0$, the only solution of $\{J_Z = 0\}$ for $X_2$ is
\begin{equation*}
X_2 = - \frac{\beta_2 X_3 + \beta_3 X_1}{\beta_1 - \beta_2 - \beta_3}. \\
\end{equation*}
(For simplicity we assume $\beta_1 - \beta_2 - \beta_3 \neq 0$.)
Then, in this limit, the resulting $J$ functions are given by
\begin{align*}
J_1&= - \frac{q_3}{X_1} + \frac{( A_1 X_1 + A_2 X_3)( E_1 X_1 + E_2 X_3)}{X_1} , \\
J_3 &= - \frac{q_2}{X_3} + \frac{( \Gamma_1 X_1 + \Gamma_2 X_3)( \Delta_1 X_1 + \Delta_2 X_3)}{X_3}, \\
\end{align*}
where
\begin{align*}
A_1 &= \frac{ \beta_3 (\alpha_2 - \alpha_1 ) + \alpha_3 (\beta_1 - \beta_2 )}{\beta_1 - \beta_2 - \beta_3}, 
\quad
A_2 = \frac{\alpha_2 ( \beta_1 - \beta_3 ) + \beta_2 (\alpha_3 - \alpha_1 )}{\beta_1 - \beta_2 - \beta_3},
\\
E_1 &= \frac{ \beta_3 (\epsilon_2 - \epsilon_1 ) + \epsilon_3 (\beta_1 - \beta_2 )}{\beta_1 - \beta_2 - \beta_3}, 
\quad
E_2 = \frac{\epsilon_2 ( \beta_1 - \beta_3 ) + \beta_2 (\epsilon_3 - \epsilon_1 )}{\beta_1 - \beta_2 - \beta_3}, 
\\
\Gamma_1 &= \frac{ \beta_3 (\gamma_2 - \gamma_1 ) + \gamma_3 (\beta_1 - \beta_2 )}{\beta_1 - \beta_2 - \beta_3}, 
\quad
\Gamma_2 = \frac{\gamma_2 ( \beta_1 - \beta_3 ) + \beta_2 (\gamma_3 - \gamma_1 )}{\beta_1 - \beta_2 - \beta_3}, 
\\
\Delta_1 &= \frac{ \beta_3 (\delta_2 - \delta_1 ) + \delta_3 (\beta_1 - \beta_2 )}{\beta_1 - \beta_2 - \beta_3},
\quad
\Delta_2 = \frac{\delta_2 ( \beta_1 - \beta_3 ) + \beta_2 (\delta_3 - \delta_1 )}{\beta_1 - \beta_2 - \beta_3}.
\end{align*}

The $J$'s above are equivalent to (\ref{eq:p1p1_1}), (\ref{eq:p1p1_2}) in 
the mirror to the $A/2$ model on $\mathbb{P}^1 \times \mathbb{P}^1$, if we take
\[
A = 
\begin{bmatrix}
A_1 & 0 \\
0 & E_1
\end{bmatrix}, 
\quad
B = 
\begin{bmatrix}
A_2 & 0 \\
0 & E_2
\end{bmatrix},
\]
\[
C = 
\begin{bmatrix}
\Gamma_1 & 0 \\
0 & \Delta_1
\end{bmatrix}, 
\quad
D = 
\begin{bmatrix}
\Gamma_2 & 0 \\
0 & \Delta_2
\end{bmatrix}.
\]

In principle one could also
similarly analyze the mirror of the blowdown 
$dP_2 \to \mathbb{P}^1 \times \mathbb{P}^1$ in the same limit $q_1 \to 0$ in
terms of the $J$ functions (\ref{eq:dp2-alt-1-1})-(\ref{eq:dp2-alt-1-4}), but
we will not do so here.

\subsection{The third del Pezzo surface, $dP_3$}

\subsubsection{Review of the $(2,2)$ mirror}

In this section, we will consider the last toric del Pezzo $dP_3$,
 which can be described by a fan
with edges $(1,0), (0,1), (-1, -1), (1,1), (-1,0), (0,-1)$. 
The corresponding GLSM has six chiral superfields $\phi_i, i=1, \dots, 6$ 
which are charged under the gauge group $U(1)^4$ as follows:
\[
\begin{tabular}{cccccc}
(1,0) & (-1,-1) & (0,1) & (0,-1) & (-1,0) & (1, 1) \\
\hline
1 & 1 & 1 & 0 & 0 & 0 \\
0 & 0 & 1 & 1 & 0 & 0 \\
1 & 0 & 0 & 0 & 1 & 0 \\
0 & 1 & 0 & 0 & 0 & 1 \\
\end{tabular}
\]

Following section~\ref{sect:review:22},
the superpotential of the (2,2) mirror is given by,
\begin{displaymath}
W = \sum_{i=1}^6 \exp (  Y_i) = \sum_{i=1}^6 X_i
\end{displaymath}
for $X_i = \exp(Y_i)$,
with constraints:
\begin{displaymath}
Y_1 + Y_2 + Y_3 = r_1, \quad Y_3 + Y_4 = r_2, \quad Y_1 + Y_5 = r_3, 
\quad Y_2 + Y_6 = r_4.
\end{displaymath}

Eliminating $X_2$ and $X_4$ via two of the constraints above,
and
introducing two Lagrange multipliers $Z_1$ and $Z_2$ to implement the
remaining constraints, 
the superpotential can be written as
\begin{equation}
W= X_1 + X_3 + X_5 + X_6 + \frac{q_1}{X_1 X_3} + \frac{q_2}{X_3} 
+ Z_1 \left( 1 - \frac{q_3}{X_1 X_5} \right) 
+ Z_2 \left(1 - \frac{q_4}{q_1} \frac{X_1 X_3}{X_6}\right).  \label{eq:dp3-22}
\end{equation}
The vacua solve the following algebraic equations:
\begin{align*}
X_1 \partial_1 W &=  X_1 - \frac{q_1}{X_1 X_3} + Z_1 \frac{q_3}{X_1 X_5} 
- Z_2 \frac{q_4}{q_1} \frac{X_1 X_3}{X_6} = 0, \\
X_3 \partial_3 W &=  X_3 - \frac{q_2}{X_3} - \frac{q_1}{X_1 X_3} 
- Z_2 \frac{q_4}{q_1} \frac{X_1 X_3}{X_6} = 0, \\
X_5 \partial_5 W &=  X_5 + Z_1 \frac{q_3}{X_1 X_5} = 0, \\
X_6 \partial_6 W &=  X_6 + Z_2 \frac{q_4}{q_1} \frac{X_1 X_3}{X_6} = 0, \\
\partial_{Z_1} W &= 1 - \frac{q_3}{X_1 X_5} = 0, \\
\partial_{Z_2} W &= 1- \frac{q_4}{q_1} \frac{X_1 X_3}{X_6} = 0.
\end{align*}

The quantum cohomology relations are
\begin{align*}
 (\psi_1 + \psi_3) (\psi_1 + \psi_4) (\psi_1 + \psi_2) &= q_1, \\
 (\psi_1 + \psi_2) \psi_2 &= q_2, \\
 (\psi_1 + \psi_3) \psi_3 &= q_3, \\
 (\psi_1 + \psi_4) \psi_4 &= q_4. \\
\end{align*}
One can check that these quantum cohomology ring relations are satisfied on the space of vacua of the Toda theory after identifying 
\begin{equation*}
X_1 \sim \psi_1 + \psi_3, \quad X_3 \sim \psi_1 + \psi_2, \quad  X_5 \sim \psi_3, \quad X_6 \sim \psi_4.
\end{equation*}
One can also check that all the correlation functions match those of 
the $A$-twisted theory on $dP_3$.

\subsubsection{$(0,2)$ deformations and proposed (0,2) mirrors}

To describe the $(0,2)$ deformation of $dP_3$, we will need 24 complex parameters $\alpha_i, \beta_j, \gamma_k, \delta_l, \epsilon_m, \zeta_n, i,j,k,l,m,n= 1 \dots 4$. Those parameters define a deformation $\mathcal{E}$ of the tangent bundle as follows,
\begin{align*}
0 \: \longrightarrow {\cal O}^3 \: \xrightarrow{E} \; {\cal O}(1,0,1,0) \oplus & {\cal O}(1,0,0,1) \oplus {\cal O}(1,1,0,0) \oplus {\cal O}(0,1,0,0)    \\
   & \quad \oplus {\cal O}(0,0,1,0) \oplus {\cal O}(0,0,0,1) \; \longrightarrow \: {\cal E} \: \longrightarrow 0,
\end{align*}
where $E$ is defined by:
\begin{displaymath}
E =
\begin{bmatrix}
\alpha_1 s_1 & \alpha_2 s_2 & \alpha_3 s_3 & \alpha_4 s_4 \\
\beta_1 s_1 & \beta_2 s_2 & \beta_3 s_3 & \beta_4 s_4 \\
\gamma_1 s_1 & \gamma_2 s_2 & \gamma_3 s_3 & \gamma_4 s_4 \\
\delta_1 s_1 & \delta_2 s_2 & \delta_3 s_3 & \delta_4 s_4 \\
\epsilon_1 s_1 & \epsilon_2 s_2 & \epsilon_3 s_3 & \epsilon_4 s_4 \\
\zeta_1 s_2 & \zeta_2 s_2 & \zeta_3 s_3 & \zeta_4 s_4 \\
\end{bmatrix} ,
\end{displaymath}
for $s_i$ the chiral superfields of the GLSM.
The (2,2) locus is given by the special case
\begin{align*}
& \alpha_1 = 1, \alpha_2 = 0, \alpha_3 = 1, \alpha_4 = 0, \\
& \beta_1 = 1, \beta_2 = \beta_3 = 0, \beta_4 = 1, \\
& \gamma_1 = \gamma_2 = 1, \gamma_3 = \gamma_4 = 0, \\
& \delta_1 = 0, \delta_2 = 1, \delta_3 = \delta_4 = 0, \\
& \epsilon_1 = \epsilon_2 = 0, \epsilon_3 = 1, \epsilon_4 =0, \\
& \zeta_1 = \zeta_2 = \zeta_3 = 0, \zeta_4 = 1.\\
\end{align*}
If we define:
\begin{align*}
& Q_{(1)} = \sum_{i=1}^4 \alpha_i \psi_i, \quad Q_{(2)} = \sum_{i=1}^4 \beta_i \psi_i, \quad Q_{(3)} = \sum_{i=1}^4 \gamma_i \psi_i,  \\
& Q_{(4)} = \sum_{i=1}^4 \delta_i \psi_i, \quad Q_{(5)} = \sum_{i=1}^4 \epsilon_i \psi_i, \quad Q_{(6)} = \sum_{i=1}^4 \zeta_i \psi_i,
\end{align*}
then the quantum sheaf cohomology ring relations are
\begin{align}
Q_{(1)} Q_{(2)} Q_{(3)} &= q_1, \label{eq: dp3_q1} \\
Q_{(3)} Q_{(4)} &= q_2, \\
Q_{(1)} Q_{(5)} &= q_3, \\
Q_{(2)} Q_{(6}) &= q_4, \label{eq: dp3_q4}
\end{align}
which reduce to the
ordinary quantum cohomology ring relations on the $(2,2)$ locus.

Our proposal for the (0,2) mirror of the $A/2$-twisted theory on $dP_3$ 
with a deformation of the tangent bundle is defined by the following
six $J$ functions:
\begin{align*}
J_1 &= - \frac{q_1}{X_1 (\gamma \cdot X)} 
+ Z_1 \frac{q_3}{X_1 (\epsilon \cdot X)} 
- Z_2 \frac{q_4}{q_1} \frac{ (\alpha \cdot X) (\gamma \cdot X) }{ (\zeta \cdot X) } 
+ \frac{ (\alpha \cdot X) (\epsilon \cdot X)}{X_1}  \nonumber \\
& \quad - (\zeta \cdot X) + \frac{ (\alpha \cdot X) (\beta \cdot X)}{X_1},  \\
J_3 &= - \frac{q_2}{X_3} - \frac{q_1}{X_1 (\gamma \cdot X)}  
- Z_2 \frac{q_4}{q_1} \frac{ (\alpha \cdot X) (\gamma \cdot X) }{ (\zeta \cdot X) } 
+ \frac{ (\alpha \cdot X) (\beta \cdot X)}{X_1} 
- (\zeta \cdot X)  \nonumber \\
&\quad + \frac{ (\gamma \cdot X) (\delta \cdot X)}{X_3} , \\
J_5 &=  (\epsilon \cdot X) + Z_1 \frac{q_3}{ (\alpha \cdot X) (\epsilon \cdot X)}, \\
J_6 &=  (\zeta \cdot X) + Z_2 \frac{q_4}{q_1} \frac{ (\alpha \cdot X) (\gamma \cdot X) }{ (\zeta \cdot X) } , \\
J_{Z1} &= \frac{ (\epsilon \cdot X) }{X_5} - \frac{q_3}{X_5 (\alpha \cdot X)}, \\
J_{Z2} &= \frac{ (\zeta \cdot X) }{X_6} - \frac{q_4}{q_1} \frac{ (\alpha \cdot X) (\gamma \cdot X) }{ X_6 },
\end{align*}
where
\begin{align*}
(\alpha \cdot X) &= \alpha_1 (X_1 - X_5) + \alpha_2 (-X_1 + X_3 + X_5) + \alpha_3 X_5 + \alpha_4 X_6, \\
& \; \; \vdots  \\
(\zeta \cdot X) &= \zeta_1 (X_1 - X_5) + \zeta_2 (-X_1 + X_3 + X_5) + \zeta_3 X_5 + \zeta_4 X_6.
\end{align*} 
(Because the $J$'s have poles away from origins, we interpret the resulting
action in a low-energy effective field theory sense, as discussed in the
introduction.)

On the (2,2) locus, the above $J$ functions reduce to derivatives of
the (2,2) superpotential (\ref{eq:dp3-22}) as expected. 
It is also easy to check that this theory respects the $U(1)_V$ symmetry,
but $U(1)_A$ is broken classically.
One can also show that on the space of vacua all quantum sheaf cohomology ring 
relations (\ref{eq: dp3_q1})-(\ref{eq: dp3_q4}) are satisfied after identifying 
\begin{equation*}
X_1 = \psi_1 + \psi_3, \quad X_3 = \psi_1 + \psi_2, \quad X_5 =  \psi_3, \quad X_6 = \psi_4.
\end{equation*}

In all other examples in this paper, we have checked that all of
the genus zero correlation functions of the 
proposed $B/2$-twisted Landau-Ginzburg mirror match those of 
the original $A/2$ theory, for all deformations.  However, for
$dP_3$, we have only checked that the genus zero 
correlation functions match in several
families of deformation parameters, described below:
\begin{enumerate}
\item families parametrized by $\alpha_i$, $\beta_i$, $\gamma_i$,
$\delta_i$, $\epsilon_i$, $\zeta_i$, for fixed $i \in \{1, \cdots, 4\}$,
and other parameters set to their (2,2) locus values,
\item \begin{align*}
& \alpha_2 = \alpha_4 = 0, \beta_2 = \beta_3 = 0, \\
& \gamma_3 = \gamma_4 = 0, \delta_1 = \delta_3 = \delta_4 = 0, \\
& \epsilon_1 = \epsilon_2 = \epsilon_4 = 0, 
\zeta_1 = \zeta_2 = \zeta_3 = 0,
\end{align*}
for a family parametrized by $\alpha_{1, 3}$, $\beta_{1, 4}$, $\gamma_{1, 2}$,
$\delta_2$, $\epsilon_3$, $\zeta_4$,
 \item \begin{align*}
& \alpha_1 = 1, \alpha_2 = 0, \alpha_3 = 1, \alpha_4 = 0, \\
& \beta_1 = 1, \beta_2 = \beta_3 = 0, \beta_4 = 1,
\end{align*}
for a family parametrized by
$\gamma_{1-4}$, $\delta_{1-4}$, $\epsilon_{1-4}$, $\zeta_{1-4}$,
\item \begin{align*}
& \gamma_1 = \gamma_2 = 1, \gamma_3 = \gamma_4 = 0, \\
& \epsilon_1 = \epsilon_2 = 0, \epsilon_3 = 1, \epsilon_4 =0,
\end{align*}
for a family parametrized by
$\alpha_{1-4}$, $\beta_{1-4}$, $\delta_{1-4}$, $\zeta_{1-4}$.
\end{enumerate}
For each of the families of deformation parameters above, 
we have checked that all of the genus zero correlation functions of
the proposed
$B/2$-twisted Landau-Ginzburg model match those
of the original $A/2$-twisted theory.

\subsubsection{Consistency check:  mirrors of blowdowns to $dP_2$}

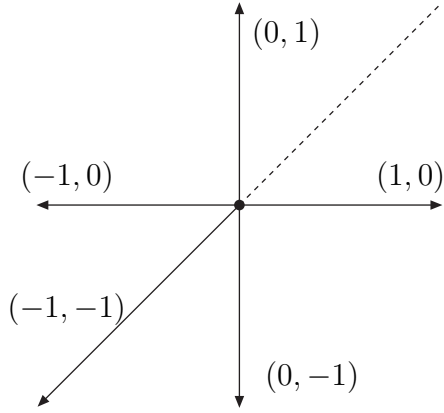
\begin{figure}[htb]
\begin{center}
\begin{picture}(150,150)
\LongArrow(75,75)(150,75)  \Text(140,80)[b]{$(1,0)$}
\LongArrow(75,75)(75,150)  \Text(80,140)[l]{$(0,1)$}
\LongArrow(75,75)(0,0)      \Text(10,30)[b]{$(-1,-1)$}
\LongArrow(75,75)(0,75)     \Text(10,80)[b]{$(-1,0)$}
\LongArrow(75,75)(75,0)    \Text(85,10)[l]{$(0,-1)$}
\DashLine(75,75)(150,150){2}
\Vertex(75,75){2}
\end{picture} 
\end{center}
\caption{A toric fan for $dP_2$ can be obtained by removing the edge $(1,1)$
from the toric fan for $dP_3$.
\label{fig:dp3-dp2}}
\end{figure}

In this section we will describe the mirror of the blowdown
$dP_3 \rightarrow dP_2$, verifying that the blowdown of the mirror
is the mirror of the blowdown.  This can be represented torically
by removing the edge $(1,1)$ from the toric fan previously discussed
for $dP_3$, as shown in figure~\ref{fig:dp3-dp2}.
(As before, since all of the varieties in question are Fano, the UV phases
of the GLSMs correspond to the geometries described here.)

The (2,2) mirror of this blowdown is given by applying the limit $q_4 \to 0$ to the superpotential (\ref{eq:dp3-22}) after integrate out one of the Lagrange multiplier $Z_2$. Following the same procedure for (0,2) mirror dual, one first integrate out the Lagrange multiplier $Z_2$ and obtain two constraints:
\begin{align*}
J_6 &=   (\zeta \cdot X) 
+ Z_2 \frac{q_4}{q_1} \frac{ (\alpha \cdot X) (\gamma \cdot X) }{ (\zeta \cdot X) } = 0, \\
J_{Z2} &=  \frac{ (\zeta \cdot X) }{X_6}
 - \frac{q_4}{q_1} \frac{ (\alpha \cdot X) (\gamma \cdot X) }{ X_6 } = 0.
\end{align*}

In the limit $q_4 \to 0$, the constraints above imply $\zeta \cdot X=0$ and
$Z_2=0$, or
for $\zeta_4 \neq 0$ (which we assume for simplicity),
\begin{displaymath}
X_6 =  -\zeta_4^{-1} \left( \zeta_1 (X_1 - X_5) + \zeta_2 (-X_1 + X_3 + X_5) + \zeta_3 X_5 \right), 
\quad
Z_2 = 0.
\end{displaymath}

Plugging those constraints into the $J$s, one gets
\begin{align*}
J_1 &= - \frac{q_1}{X_1 (\gamma \cdot X)} 
+ Z_1 \frac{q_3}{X_1 (\epsilon \cdot X)}  
+ \frac{ (\alpha \cdot X) (\epsilon \cdot X)}{X_1} 
+ \frac{ (\alpha \cdot X) (\beta \cdot X)}{X_1},  \\
J_3 &= - \frac{q_2}{X_3} - \frac{q_1}{X_1 (\gamma \cdot X)} 
 + \frac{ (\alpha \cdot X) (\beta \cdot X)}{X_1}
 + \frac{ (\gamma \cdot X) (\delta \cdot X)}{X_3} , \\
J_5 &=  (\epsilon \cdot X) 
+ Z_1 \frac{q_3}{ (\alpha \cdot X) (\epsilon \cdot X)}, \\
J_{Z1} &= \frac{ (\epsilon \cdot X) }{X_5} - \frac{q_3}{X_5 (\alpha \cdot X)}, 
\end{align*}
where
\begin{align*}
(\alpha \cdot X) &= A_1 (X_1 - X_5) + A_2 (-X_1 + X_3 + X_5) + A_3 X_5, \\
(\beta \cdot X) &= B_1 (X_1 - X_5) + B_2 (-X_1 + X_3 + X_5) + B_3 X_5, \\
(\gamma \cdot X) &= G_1 (X_1 - X_5) + G_2 (-X_1 + X_3 + X_5) + G_3 X_5, \\
(\delta \cdot X) &= D_1 (X_1 - X_5) + D_2 (-X_1 + X_3 + X_5) + D_3 X_5, \\
(\epsilon \cdot X) &= E_1(X_1 - X_5) + E_2 (-X_1 + X_3 + X_5) + E_3 X_5, \\
\end{align*}
with 
\begin{align*}
&A_1 = \alpha_1 - \alpha_4 \zeta_1 \zeta_4^{-1}, \quad A_2 = \alpha_2 - \alpha_4 \zeta_2 \zeta_4^{-1}, \quad A_3 = \alpha_3 - \alpha_4 \zeta_3  \zeta_4^{-1}, \\
&B_1 = \beta_1 - \beta_4 \zeta_1 \zeta_4^{-1}, \quad B_2 = \beta_2 - \beta_4 \zeta_2 \zeta_4^{-1}, \quad B_3 = \beta_3 - \beta_4 \zeta_3  \zeta_4^{-1}, \\
&G_1 = \gamma_1 - \gamma_4 \zeta_1 \zeta_4^{-1}, \quad G_2 = \gamma_2 - \gamma_4 \zeta_2 \zeta_4^{-1}, \quad G_3 = \gamma_3 - \gamma_4 \zeta_3  \zeta_4^{-1}, \\
&D_1 = \delta_1 - \delta_4 \zeta_1 \zeta_4^{-1}, \quad D_2 = \delta_2 - \delta_4 \zeta_2 \zeta_4^{-1}, \quad D_3 = \delta_3 - \delta_4 \zeta_3  \zeta_4^{-1}, \\
&E_1 = \epsilon_1 - \epsilon_4 \zeta_1 \zeta_4^{-1}, \quad E_2 = \epsilon_2 - \epsilon_4 \zeta_2 \zeta_4^{-1}, \quad E_3 = \epsilon_3 - \epsilon_4 \zeta_3  \zeta_4^{-1}.
\end{align*}
The resulting $J$ functions are the same as those of $dP_2$ in
(\ref{eq:dp2-alt-1-1})-(\ref{eq:dp2-alt-1-4}) 
with parameters 
\begin{align*}
\alpha_1 = A_1, \quad \alpha_2 = A_2, \quad \alpha_3 = A_3, \\
\beta_1 = B_1, \quad \beta_2 = B_2, \quad \beta_3 = B_3, \\
\gamma_1 = B_1, \quad \gamma_2 = B_2, \quad \gamma_3 = B_3, \\
\delta_1 = D_1, \quad \delta_2 = D_2, \quad \delta_3 = D_3, \\
\epsilon_1 = E_1, \quad \epsilon_2 = E_2, \quad \epsilon_3 = E_3.
\end{align*}

\section{Hirzebruch surfaces} \label{sec: hirzebruch}

\subsection{Review of the (2,2) mirror}
 
We will first review the construction of Hirzebruch surfaces $\mathbb{F}_n$ 
and their (2,2) Toda duals along with the ordinary quantum cohomology relations.

Recall a 
Hirzebruch surface $\mathbb{F}_n$  is a toric variety which can be described
by the fan with edges $(1,0)$, $(0,1)$, $(0,-1)$, $(-1,-n)$.  
The corresponding gauged linear sigma model has four chiral superfields $\phi_1$, 
$\phi_2$, $\phi_3$, $\phi_4$ which are charged under the gauge group 
$U(1)^2$ as follows
\begin{center}
\begin{tabular}{ l c c r }
 $u$ & $v$ & $s$ & $t$ \\
 \hline
 1 & 1 & n & 0 \\
 0 & 0 & 1 & 1
\end{tabular}
\end{center}

Now, for $n>1$, Hirzebruch surfaces ${\mathbb F}_n$ are not Fano.
This fact manifests itself in the RG flow:  en route to the IR, the GLSM
for a Hirzebruch surface will enter a different phase, and describe a
different geometry.  Nevertheless, we expect them to flow in the IR to
a discrete set of isolated vacua, and so one can reasonably expect a
Toda-type mirror, despite the fact that they are not Fano. 
On the other hand, for the same reasons,
for $n>2$ the mirrors we describe should not be
interpreted as mirrors to Hirzebruch surfaces {\it per se} but rather
to different phases of the same GLSM, specifically to geometries
${\mathbb P}^2_{[1,1,n]}$ which appear as the UV phases of the same GLSMs.
In any event, for simplicity,
we will speak loosely of `mirrors to Hirzebruch surfaces' with the
understanding that we are speaking of mirrors to GLSMs, and the actual
geometries being mirrored are the UV phases, which for $n>2$ will not
in fact be Hirzebruch surfaces.

Following the methods of \cite{Hori:2000kt} (with the non-Fano caveat
above),
the mirror is a Landau-Ginzburg model with superpotential
\begin{displaymath}
W = \exp( Y_1) + \exp( Y_2) + \exp( Y_3) + \exp( Y_4), 
\end{displaymath}
with constraints
\begin{displaymath}
Y_1 + Y_2 + n Y_3 = r_1, \quad Y_3 + Y_4 = r_2. 
\end{displaymath}
We can then use those two constraints to write the result in terms of only
$Y_1$, $Y_3$,
yielding \cite{Hori:2000kt}[equ'n (5.19)]
\begin{displaymath}
W = X_1 + X_3 + \frac{q_2}{X_3} + \frac{ q_1}{ X_1 X_3^{n}}, 
\end{displaymath}
where we defined $X_1 = \exp( Y_1)$, $X_3 = \exp( Y_3)$, 
$q_1 = \exp( r_1)$ and $q_2 = \exp( r_2)$.

As discussed in e.g. \cite{Morrison:1994fr},
the quantum cohomology ring relations are given by
\begin{displaymath}
\psi^2 (n \psi + \tilde{\psi})^n = q_1, \quad \tilde{\psi} (n \psi + \tilde{\psi}) = q_2.
\end{displaymath}
To translate to the present case, we use the dictionary
\begin{equation}
\psi \sim X_1, \quad n \psi + \tilde{\psi} \sim X_3, \label{eq:hir-X-psi}
\end{equation}
so on the space of vacua, we expect that the fields $X_1, X_3$ of the 
mirror should obey
\begin{equation}  \label{eq:hirz-22-vacua}
X_1^2 X_3^n = q_1, \quad (X_3 - n X_1) X_3 = q_2.
\end{equation}
Vacua are computed by taking derivatives of the superpotential
with respect to $\ln X_1, \ln X_3$.  Doing so one finds that the vacua are
defined by
\begin{align}
X_1 - q_1 X_1^{-1} X_3^{-n} = 0 \label{eq:1} , \\
X_3 - q_2 X_3^{-1} - n q_1 X_1^{-1} X_3^{-n} = 0.
\end{align}
After some algebra one can show that these conditions for vacua imply
conditions~(\ref{eq:hirz-22-vacua}), as expected.

For geometries previously described, we have given alternative
mirrors, and this is no exception.  Solving the original constraints
for $Y_1$ and $Y_3$, and defining
$X_1 = \exp ( Y_1)$ and $X_3 = \exp ( Y_3)$,
we get an alternative expression for the superpotential defining the
(2,2) mirror:
\begin{displaymath}
W = X_1 + q_1 q_2^{-n} X_1^{-1} X_4^n + X_4 + q_2 X_4^{-1}.
\end{displaymath}
This expression is related to the one above by the field redefinition
\begin{displaymath}
X_4 \: = \: \frac{q_2}{X_3}.
\end{displaymath}

\subsection{(0,2) deformations and proposed (0,2) mirrors} \label{sec:fn-02}

In this section, we will give a proposal for the mirror
$(0,2)$ Landau-Ginzburg model to an $A/2$-twisted nonlinear sigma model on
a Hirzebruch surface with a deformation of the tangent bundle,
which we will check by matching correlation functions.

The $(0,2)$ deformations of a Hirzebruch surface $\mathbb{F}_n$
are defined by a pair of $2 \times 2$ matrices $A,B$,
and complex numbers $\gamma_1, \gamma_2, \alpha_1, \alpha_2$,
that define a deformation $\mathcal{E}$ of the tangent bundle
\[
0 \: \longrightarrow \: {\cal O}^{\oplus 2} \: \xrightarrow{E} \: {\cal O}(1,0)^{\oplus 2} \oplus {\cal O}(n,1) \oplus {\cal O}(0,1) \: \longrightarrow {\cal E} \: \longrightarrow 0 ,
\]
where $E$ is
\begin{displaymath}
E = 
\begin{bmatrix}
Ax & Bx \\
\gamma_1 s & \gamma_2 s \\
\alpha_1 t & \alpha_2 t
\end{bmatrix},
\end{displaymath}
with
\begin{displaymath}
x= 
\begin{bmatrix}
u \\
v
\end{bmatrix}.
\end{displaymath}
The $(2,2)$ locus is given by the special case
\begin{displaymath}
A = I, \quad B = 0, \quad \gamma_1 = n, \quad \gamma_2 = 1, \quad \alpha_1 = 0, \quad \alpha_2 = 1.
\end{displaymath}
If we define
\begin{displaymath}
Q_{(k)} = \det (\psi A + \tilde{\psi} B), \quad Q_{(s)} = \psi \gamma_1 + \tilde{\psi} \gamma_2, \quad Q_{(t)} = \psi \alpha_1 + \tilde{\psi} \alpha_2,
\end{displaymath}
then the quantum sheaf cohomology ring relations are given by \cite{Melnikov:2012hk}
\begin{displaymath}
Q_{(k)} Q_{(s)}^n = q_1, \quad Q_{(s)} Q_{(t)} = q_2.
\end{displaymath}

Our proposal for the $(0,2)$ Toda mirror of the $A/2$ model on
${\mathbb F}_n$ with a deformation of the tangent bundle is
defined by
\begin{align}
J_1 &=  a X_1 + \mu_{AB} (X_3 - n X_1) + b \frac{(X_3 - n X_1)^2}{X_1}  \nonumber \\
& \qquad \qquad \qquad \qquad \qquad \qquad - q_1 X_1^{-1} \left( \gamma_1 X_1 + \gamma_2 (X_3 - n X_1) \right)^{-n}, \label{eq:hir-02-1}\\
J_2 &= n \left(  a X_1 + \mu_{AB} (X_3 - n X_1) 
+ b \frac{(X_3 - n X_1)^2}{X_1} \right) \nonumber  \\
& \qquad \qquad \qquad  - \frac{ n q_1 }{ X_1 
\left( \gamma_1 X_1 + \gamma_2 (X_3 - n X_1) \right)^{n}  }
- \frac{q_2}{X_3} \nonumber \\
&\qquad \qquad \qquad \qquad + \frac{
\left( \gamma_1 X_1 + \gamma_2 (X_3 - n X_1) \right) \left( \alpha_1 X_1 + \alpha_2 (X_3 - n X_1) \right) }{X_3}. \label{eq:hir-02-2}
\end{align}
(Because the $J$'s have poles away from origins, we interpret the resulting
action in a low-energy effective field theory sense, as discussed in the
introduction.)
In the expression above,
we used the same notation as in our description of the $A/2$ theory, namely
\begin{displaymath}
a = \det A, \quad b = \det B, \quad \mu_{AB} = \det (A+B) - \det A - \det B.
\end{displaymath}

We will check our proposal by arguing that $A/2$ model
correlation functions will match those of the $B/2$-twisted mirror
Landau-Ginzburg theory given above.  Before doing so,
let us first make a few elementary observations. As one might expect, our proposal reduces to the $(2,2)$ Toda dual when $\mathcal{E} = TX$,
corresponding to $A = I, B = 0, \gamma_1 = n, \gamma_2 = 1, \alpha_1 = 0, \alpha_2 = 1$,
as in this case each $J_i$ becomes the derivative of the
(2,2) superpotential with respect to $Y_1=\ln X_1$ and $Y_3=\ln X_3$. 
As another consistency check, 
one can show $X_1, X_3$ satisfy the quantum sheaf cohomology relations on the 
space of vacua.  Specifically, the vacua are defined by
$J_1=0$, $J_2=0$, which imply
\begin{align*}
\det \left( A X_1 + B (X_3 - n X_1) \right) \left(\gamma_1 X_1 + \gamma_2 (X_3 - n X_1) \right)^n = q_1, \\ \left( \alpha_1 X_1 + \alpha_2 (X_3 - n X_1) \right) \left( \gamma_1 X_1 + \gamma_2 (X_3 - n X_1) \right)= q_2.
\end{align*}
With the correspondence (\ref{eq:hir-X-psi}), 
it is straightforward to show that the quantum sheaf cohomology relations
are satisfied on the vacua.

As another consistency check, we observe that this naturally specializes to
results obtained in \cite{Chen:2016tdd} and reviewed in
section~\ref{sect:rev02-products} for the mirror to the $A/2$ model on
${\mathbb P}^1 \times {\mathbb P}^1$ with a deformation of the tangent bundle.
If we take $n=0$, then the resulting Hirzebruch surface with tangent bundle
deformation corresponds to ${\mathbb P}^1 \times {\mathbb P}^1$ with
\begin{displaymath}
C \: = \: \left[ \begin{array}{cc}
\gamma_1 & 0 \\
0 & \alpha_1 \end{array} \right], \: \: \:
D \: = \: \left[ \begin{array}{cc}
\gamma_2 & 0 \\
0 & \alpha_2 \end{array} \right],
\end{displaymath}
so that, after simplification,
\begin{eqnarray}
J_1 & = & a X_1 + \mu_{AB} X_2 + b \frac{X_2^2}{X_1} - \frac{q_1}{X_1}, \label{eq:p1xp1-1} \\
J_2 & = & \alpha_1 \gamma_1 \frac{X_1^2}{X_2} +
(\alpha_1 \gamma_2 + \alpha_2 \gamma_1) X_1 + \alpha_2 \gamma_2 X_2 -
\frac{q_2}{X_2}. \label{eq:p1xp1-2}
\end{eqnarray}
One can easily observe that the $J$ functions above (\ref{eq:p1xp1-1}), 
(\ref{eq:p1xp1-2}) are the same as (\ref{eq:hir-02-1}), 
(\ref{eq:hir-02-2}) after setting $n \mapsto 0$. 

Similarly, for the case $n=1$, the mirror here matches
the mirror to $dP_1 = {\mathbb F}_1$ described previously
in section~\ref{sect:dp1:02mirror}.

We have checked that all genus zero correlation functions in this proposed
(0,2) mirror match those of the original $A/2$-twisted theory, following the
arguments outlined in section~\ref{sect:dp1:02mirror}.

So far we have presented a (0,2) mirror proposal that reduces on the (2,2)
locus to the first expression for a (2,2) mirror.  As we have done for
other geometries, we next present a (0,2) mirror proposal that reduces
on the (2,2) locus to the second expression for a (2,2) mirror.
Specifically, a
proposal for a $(0,2)$ mirror of the $A/2$ model on $\mathbb{F}_n$ 
(with a deformation of the tangent bundle) that reduces on the (2,2)
locus to the model above is given by
\begin{align}
J_1 &=  \left( a X_1 + \mu_{AB} X_4 + b \frac{X_4^2}{X_1} \right) 
- \frac{q_1}{q_2^n} \frac{(\alpha_1 X_1 + \alpha_2 X_4)^n}{X_1}, \label{eq:hir-02-3} \\
J_2 &=  
  \left( \alpha_2 \gamma_2 X_4 + \gamma_1 \alpha_1 \frac{X_1^2}{X_4} 
+ \frac{q_1}{q_2^n} \frac{(\alpha_1 X_1 + \alpha_2 X_4)^n (\gamma_1 \alpha_2 + \gamma_2 \alpha_1)}{a X_1 + \mu_{AB} X_4 + b X_4^2 X_1^{-1} } 
\right) 
- \frac{q_2}{X_4}. \label{eq:hir-02-4}
\end{align}

In passing, we should mention that an alternative expression which also
has matching correlation functions and the correct (2,2) locus can be written
which has the same $J_1$ but a different $J_2$ given by
\begin{align}
J_2 = & -n  \left( a X_1 + \mu_{AB} X_4 + b \frac{X_4^2}{X_1}  
- \frac{q_1}{q_2^n} \frac{(\alpha_1 X_1 + \alpha_2 X_4)^n}{X_1}
\right)   \nonumber
\\
& +  \left( \alpha_2 \gamma_2 X_4 + \gamma_1 \alpha_1 \frac{X_1^2}{X_4} 
+ (\gamma_1 \alpha_2 + \gamma_2 \alpha_1) X_1
\right) 
- \frac{q_2}{X_4}.
\end{align}
Of course, by taking suitable field redefinitions, we can simplify
the second $J_2$ above to write it in the form
\begin{equation}
J_2 \: = \:
 \left( \alpha_2 \gamma_2 X_4 + \gamma_1 \alpha_1 \frac{X_1^2}{X_4} 
+ (\gamma_1 \alpha_2 + \gamma_2 \alpha_1) X_1
\right) 
- \frac{q_2}{X_4}.
\end{equation}
This third model, with the altered $J_2$ above,
 does not reduce to the (2,2) locus expression given 
previously, but we felt important to point out its existence.

One can check that these alternative proposals 
also reduce to the $(2,2)$ mirror, 
and $X_1$, $X_4$ satisfy the quantum sheaf cohomology relations on the space 
of vacua with identification $X_1 \sim \psi$, $X_4 \sim \tilde{\psi}$.
More importantly, all (genus zero) $A/2$ model correlation functions 
again match those of the $B/2$ Landau-Ginzburg theory given above.
Thus, this is another expression for the mirror.

Setting $n = 0$,
we also have a new expression for the $B/2$ mirror
Landau-Ginzburg theory to $\mathbb{P}^1 \times \mathbb{P}^1$,
\begin{align*}
J_1 &=   a X_1 + \mu_{AB} X_4 + b \frac{X_4^2}{X_1}  
- \frac{q_1}{X_1}, \\
J_2 &=   \alpha_2 \gamma_2 X_4 + \gamma_1 \alpha_1 \frac{X_1^2}{X_4} 
+ \frac{q_1 (\gamma_1 \alpha_2 + \gamma_2 \alpha_1)}{a X_1 + \mu_{AB} X_4 + b X_4^2 X_1^{-1} }  
- \frac{q_2}{X_4}.
\end{align*}
On the $(2,2)$ locus the $J$ functions above reduce to those of
the $(2,2)$ mirror of
$\mathbb{P}^1 \times \mathbb{P}^1$ written in $(0,2)$ language,
\begin{align*}
J_1 &=  X_1 - \frac{q_1}{X_1}, \\
J_1 &=  X_4 - \frac{q_2}{X_4}.
\end{align*}
All correlation functions of the new mirror theory
given above are the same as those given in section~(\ref{sec:fn-02}).
In both case, all correlation functions match the correlation functions of the
same one-loop effective action of the $A/2$ theory on
$\mathbb{P}^1 \times \mathbb{P}^1$.

\section{Grassmannians} \label{sec: grassmannian}

In this section we will propose a (0,2) analogue of the mirror to a 
Grassmannian proposed in \cite{Hori:2000kt}[appendix A].
As was remarked to us by one of the authors \cite{horipriv},
alternative proposals also exist in the literature, see for
example \cite{ehx,givental,r1,mr}.  In this paper, we shall only consider
(0,2) deformations of the proposal in \cite{Hori:2000kt}, and will leave
(0,2) deformations of other proposals for future work.

\subsection{(0,2) deformations} \label{deftan}

On the (2,2) locus, the Grassmannian $G(k,n)$ is described by a
two-dimensional $U(k)$ gauge theory with $n$ chirals in the
fundamental representation. We denote these chiral multiplets by
$\Phi^i_\alpha, \alpha=1,\cdots,k, i=1,\cdots,n$. 
These (2,2) chiral multiplets
decompose into (0,2) chiral multiplets $\Phi^i_\alpha
= (\phi^i_\alpha, \psi^i_{+ \alpha})$ and (0,2) Fermi multiplets
$\Lambda^i_\alpha = (\psi^i_{- \alpha}, F^i_\alpha)$, obeying
\begin{equation}\label{1}
\overline{D}_+ \Lambda^i_\alpha \: = \: \sigma^\beta_\alpha
\Phi^i_\beta .
\end{equation}

For\footnote{
In the cases $k=1, n-1,$ the Grassmannian is a projective space,
and its tangent bundle has no deformations.
} $1 < k < n-1$, there exist nontrivial
(0,2) deformations of this theory,
given explicitly as
\begin{displaymath}
\overline{D}_+ \Lambda^i_\alpha \: = \: \sigma^\beta_\alpha
\Phi^i_\beta \: + \: B^i_j ({\rm Tr} \, \sigma) \Phi^j_\alpha,
\end{displaymath}
where $B$ is an $n \times n$ matrix.
(As discussed in \cite{Guo:2015caf},
one can also rotate the first term by an $n \times n$ matrix, but that
matrix can be absorbed into field redefinitions, so for simplicity we
omit it.)
The resulting (0,2) theory
describes a deformation ${\cal E}$ of the tangent bundle, 
defined mathematically
by the short exact sequence
\begin{equation}
0 \to \mathcal{S} \otimes \mathcal{S}^* \stackrel{*}{\to}
\mathcal{V} \otimes \mathcal{S}^* \to \mathcal{E} \to 0,
\label{deformed1}
\end{equation}
where the map $*$ is given by
$\omega^{\beta}_{\alpha} \mapsto  \omega^{\beta}_{\alpha}
x^i_{\beta} + \omega^{\beta}_{\beta} B^i_j x^j_{\alpha}$.

On the Coulomb branch, the one-loop
effective $J$-functions are given by \cite{McOrist:2007kp}
\begin{equation}\label{J(0,2)}
J_a = - \log\left[ q^{-1} \det( M_a ) \right],
\end{equation}
where \[ M_a = \sigma_a I_n + \left(\sum_b \sigma_b\right) B, \] 
and $I_n$ is the $n \times n$ identity matrix.
The chiral operators are
symmetric polynomials in the $k$ fields $\sigma_a$ \cite{Guo:2015caf}.
For any such operator $\mathcal {O}$, the correlation
function of the $A/2$ twisted theory is computed by the localization
formula \cite{Closset:2015ohf}
\begin{equation}\label{corr}
\langle \mathcal {O} \rangle = \sum_{J=0} \left(\mathcal{O} \prod_{a\neq
b} (\sigma_a-\sigma_b) H^{-1}\right)
\end{equation}
where
\[
H = \det_{a,b} \left(\frac{\partial J_a}{\partial \sigma_b}\right)
\prod_a \det(M_a) .
\]

The quantum sheaf cohomology ring of this theory
is given by \cite{Guo:2015caf}
\begin{equation}  
\begin{array}{l}
{\mathbb C}\left[\sigma_{(1)}, \sigma_{(2)}, \cdots  \right] /
\left\langle D_{k+1}, D_{k+2}, \cdots, R_{(n-k+1)}, \cdots,
R_{(n-1)},
\right. \\
\hspace*{1.5in} \left. R_{(n)}+q, R_{(n+1)} + q \sigma_{(1)},
R_{(n+2)} + q \sigma_{(2)}, \cdots
 \right\rangle,
\end{array}
\end{equation}
where
\begin{align*}
D_m & =  \det\left(\sigma_{(1+j-i)} \right)_{1 \leq i,j \leq m}, \\
R_{(r)} & =  \sum_{i=0}^{ {\rm min}(r,n) } I_i \sigma_{(r-i)} \sigma_{(1)}^i,
\end{align*}
and where $I_i$ are the coefficients of the characteristic polynomial of
$B$, defined by
\begin{displaymath}
\det(\lambda I + B) \: = \: \sum_{i=0}^n I_{n-i} \lambda^i.
\end{displaymath}
(For example, $I_0=1$, independent of $B$, $I_1 = {\rm tr}\, B$,
$I_n = \det B$.)

\subsection{(0,2) mirror}

In appendix A of \cite{Hori:2000kt}, a conjecture was made 
for the mirror to an $A$-twisted GLSM for a Grassmannian.  The proposed mirror
was the ``Weyl-group-invariant part'' of
a Landau-Ginzburg model with the superpotential
\cite{Hori:2000kt}[equ'n (A.1)]:
\[
W = \sum_i \Sigma_i (Q_i^{\alpha} Y_{\alpha} - t) + \sum_{\alpha} e^{Y_{\alpha}},
\]
where $t$ is the FI parameter of the original theory,
$Y_{\alpha}$ correspond to the fundamental fields, and
$Q_i^{\alpha}$ are the charges of the fundamental fields
with respect to $U(1)^k \subset U(k)$.  Taking the Weyl-group-invariant
part meant that fundamental fields are to be written in terms of
Weyl-group-invariant combinations (not quite the same as orbifolding the
theory).

Therefore, in this section, 
we propose a (0,2) mirror of the model
introduced in section \ref{deftan}. This means the $B/2$ model of the
proposed theory should reproduce the $A/2$ chiral ring and the
correlation function \eqref{corr} of the original theory. For this
purpose, note that we can rewrite $H$ in \eqref{corr} as
\[
H = \det_{a,b}\left( - \frac{\partial \det(M_a)}{\partial \sigma_b}
\right).
\]

We propose here a (0,2) analogue of the same structure (leaving questions
about the correct physical mirror to other work).
Specifically, this proposal is built on (0,2) Landau-Ginzburg model
with chiral fields $\Sigma_a, a=1,\cdots,k$ and corresponding Fermi
fields $\Lambda_a$. The $J$ function coupling to $\Lambda_a$ is
\begin{equation}
J^a = - \det(M_a) + q.
\end{equation}
The constant $q$ is inserted to ensure that the $J$ functions of the
two theories have the same zero set. Given an operator defined by a
symmetric polynomial in the $\sigma$'s, the $B/2$ correlation function
of this dual theory is
\[
\langle \mathcal {O} \rangle = \sum_{J=0} (\mathcal{O} H^{-1}).
\]
To produce \eqref{corr} we need to define the measure of this
mirror theory to be given by
\begin{equation}
\langle \mathcal {O} \rangle = \int [D\Sigma] \prod_{a\neq b}
(\Sigma_a-\Sigma_b) \mathcal{O} e^{-S},
\end{equation}
in effect inserting factors of $\prod_{a \neq b} (\Sigma_a - \Sigma_b)$
in correlation functions, just as in the (2,2) proposal in
\cite{Hori:2000kt}[appendix A].
This clearly is not equivalent to a definition of a new QFT, but rather
is merely a (0,2) analogue of the formal structure presented in
\cite{Hori:2000kt}[appendix A].

\section{Conclusions}

In this paper, we have continued a program of trying to understand
(0,2) mirror symmetry by working out proposals for (0,2) mirrors to 
some more (non-Calabi-Yau) spaces, following up our previous work
\cite{Chen:2016tdd} on (0,2) mirrors to products of projective spaces with
tangent bundle deformations.  Specifically, we have given and checked
proposals for (0,2) mirrors to toric del Pezzo and Hirzebruch surfaces
with tangent bundle deformations, checking not only correlation functions
but also {\it e.g.} that mirrors to del Pezzos are related by blowdowns
in the fashion one would expect.

It remains for the future to find a general construction of (0,2) mirrors,
analogous to the general ansatzes in the literature for (2,2) mirrors.

\section{Acknowledgements}

We would like to thank K.~Hori, S.~Katz, I.~Melnikov,
and L.~Mihalcea for useful conversations.
E.S. was partially supported by
NSF grant PHY-1417410.

\appendix

\section{Quantum cohomology of $dP_1$}
\label{app:qc:dp1}

In this section we briefly outline standard results on the quantum
cohomology of $dP_1$, following \cite{aluffi}[section II.5].

For any $\beta = d H - \alpha E \in H_2 (dP_1, \mathbb{Z})$,
where $H$ is the pullback of the
hyperplane class of $\mathbb{P}^2$ and $E$ is
the class of the exceptional divisor (viewing $dP_1$ as the blowup of
${\mathbb P}^2$ at one point), define the Gromov-Witten invariant
\[
N_{d,\alpha} = I_{0,n_{d,\alpha},\beta}((pt)^{n_{d,\alpha}})
\]
with the expected dimension $n_{d,\alpha}=3d-1-\alpha$. Using results from
\cite{aluffi}[section II.5],
one can compute these
$N_{d,\alpha}$ recursively, and thus determine all the Gromov-Witten
invariants. For example, $N_{0,-1}=1, N_{0,-2}=0, N_{1,2}=0,
N_{1,1}=1$, from which we see
\[
\begin{split}
&I_{0,3,0}(E,E,X)=E \cdot E = -1, \\
&I_{0,3,{0,-1}}(E,E,E) = -1, \\
&I_{0,3,(1,1)}(E,E,pt) = N_{1,1} = 1,
\end{split}
\]
and all other three point Gromov-Witten invariants containing two
$E$'s vanish. Thus
\[
\begin{split}
E\ast E &= \sum_\beta \big[ I_{0,3,\beta}(E,E,X)pt +
I_{0,3,\beta}(E,E,E)(-E) + I_{0,3,\beta}(E,E,H)H \\
& \quad \quad \quad \quad \quad +
I_{0,3,\beta}(E,E,pt)X \big] q^\beta, \\
&= -pt + E q_1 + X q_0 q_1^{-1},
\end{split}
\]
where $q_1=q^E, q_0=q^H$. If we define $F=H-E$ to be the class of
the fiber and $q_2 = q^F$, then the relation can be written as
\[
E\ast E = -pt + E q_1 + X q_2.
\]
Similarly one can verify all the relations
\[
\begin{split}
E \ast F &= pt - E q_1, \\
F \ast F &= E q_1, \\
E \ast pt &= F q_2, \\
F \ast pt &= X q_1 q_2, \\
pt \ast pt &= H q_1 q_2.
\end{split}
\]

\end{document}